\documentclass[prx,a4paper,aps,onecolumn,superscriptaddress,12pt]{quantumarticle}
\usepackage{adjustbox}
\usepackage{xcolor}
\usepackage[utf8]{inputenc}
\usepackage[english]{babel}
\usepackage[T1]{fontenc}
\usepackage{amsmath}
\usepackage{hyperref}
\usepackage{setspace}
\usepackage{tikz}
\usepackage{lipsum}

\begin{document}

\title{Neuromorphic MoS$_2$ memtransistors fabricated by localised helium ion beam irradiation}
\date{\today}

\author{Jakub Jadwiszczak$^{\dagger}$}
\affiliation{Centre for Research on Adaptive Nanostructures and Nanodevices (CRANN) and Advanced Materials and Bioengineering Research (AMBER) Research Centers, Trinity College Dublin, Dublin 2, Ireland.}
\affiliation{School of Physics, Trinity College Dublin, Dublin 2, Ireland}
\affiliation{State Key Laboratory for Mesoscopic Physics, School of Physics, Peking University, Beijing 100871, People's Republic of China}
\affiliation{School of Material Science and Engineering, Nanchang University, Youxun W Rd, Xinjian Qu, Nanchang Shi, Jiangxi Sheng, People's Republic of China}

\author{Darragh Keane$^{\dagger}$}
\affiliation{Centre for Research on Adaptive Nanostructures and Nanodevices (CRANN) and Advanced Materials and Bioengineering Research (AMBER) Research Centers, Trinity College Dublin, Dublin 2, Ireland.}
\affiliation{School of Chemistry, Trinity College Dublin, Dublin 2, Ireland}

\author{Pierce Maguire}
\affiliation{Centre for Research on Adaptive Nanostructures and Nanodevices (CRANN) and Advanced Materials and Bioengineering Research (AMBER) Research Centers, Trinity College Dublin, Dublin 2, Ireland.}
\affiliation{School of Physics, Trinity College Dublin, Dublin 2, Ireland}

\author{Conor P. Cullen}
\affiliation{Centre for Research on Adaptive Nanostructures and Nanodevices (CRANN) and Advanced Materials and Bioengineering Research (AMBER) Research Centers, Trinity College Dublin, Dublin 2, Ireland.}
\affiliation{School of Chemistry, Trinity College Dublin, Dublin 2, Ireland}

\author{Yangbo Zhou}
\affiliation{School of Material Science and Engineering, Nanchang University, Youxun W Rd, Xinjian Qu, Nanchang Shi, Jiangxi Sheng, People's Republic of China}

\author{Hua-Ding Song}
\affiliation{State Key Laboratory for Mesoscopic Physics, School of Physics, Peking University, Beijing 100871, People's Republic of China}

\author{Clive Downing}
\affiliation{Centre for Research on Adaptive Nanostructures and Nanodevices (CRANN) and Advanced Materials and Bioengineering Research (AMBER) Research Centers, Trinity College Dublin, Dublin 2, Ireland.}
\affiliation{School of Chemistry, Trinity College Dublin, Dublin 2, Ireland}

\author{Daniel S. Fox}
\affiliation{Centre for Research on Adaptive Nanostructures and Nanodevices (CRANN) and Advanced Materials and Bioengineering Research (AMBER) Research Centers, Trinity College Dublin, Dublin 2, Ireland.}
\affiliation{School of Physics, Trinity College Dublin, Dublin 2, Ireland}

\author{Niall McEvoy}
\affiliation{Centre for Research on Adaptive Nanostructures and Nanodevices (CRANN) and Advanced Materials and Bioengineering Research (AMBER) Research Centers, Trinity College Dublin, Dublin 2, Ireland.}
\affiliation{School of Chemistry, Trinity College Dublin, Dublin 2, Ireland}

\author{Rui Zhu}
\affiliation{Electron Microscopy Laboratory, School of Physics, Peking University, Beijing 100871, People’s Republic of China}

\author{Jun Xu}
\affiliation{Electron Microscopy Laboratory, School of Physics, Peking University, Beijing 100871, People’s Republic of China}

\author{Georg S. Duesberg}
\affiliation{Centre for Research on Adaptive Nanostructures and Nanodevices (CRANN) and Advanced Materials and Bioengineering Research (AMBER) Research Centers, Trinity College Dublin, Dublin 2, Ireland.}
\affiliation{School of Chemistry, Trinity College Dublin, Dublin 2, Ireland}
\affiliation{Institute of Physics, EIT 2, Faculty of Electrical Engineering and Information Technology, Universit\"{a}t der Bundeswehr M\"{u}nchen, Werner-Heisenberg-Weg 39, 85577 Neubiberg,
Germany}

\author{Zhi-Min Liao}
\affiliation{State Key Laboratory for Mesoscopic Physics, School of Physics, Peking University, Beijing 100871, People's Republic of China}
\affiliation{Collaborative Innovation Center of Quantum Matter, Beijing 100871, People's Republic of China}

\author{John J. Boland}
\affiliation{Centre for Research on Adaptive Nanostructures and Nanodevices (CRANN) and Advanced Materials and Bioengineering Research (AMBER) Research Centers, Trinity College Dublin, Dublin 2, Ireland.}
\affiliation{School of Chemistry, Trinity College Dublin, Dublin 2, Ireland}

\author{Hongzhou Zhang}
\affiliation{Centre for Research on Adaptive Nanostructures and Nanodevices (CRANN) and Advanced Materials and Bioengineering Research (AMBER) Research Centers, Trinity College Dublin, Dublin 2, Ireland.}
\affiliation{School of Physics, Trinity College Dublin, Dublin 2, Ireland}
\email{hozhang@tcd.ie}

\maketitle

\onecolumngrid
\begin{abstract}
Two-dimensional layered semiconductors have recently emerged as attractive building blocks for next-generation low-power non-volatile memories. However, challenges remain in the controllable sub-micron fabrication of bipolar resistively switching circuit components from these novel materials. Here we report on the scalable experimental realisation of lateral on-dielectric memtransistors from monolayer single-crystal  molybdenum disulfide (MoS$_2$) utilising a focused helium ion beam. Site-specific irradiation with the probe of a helium ion microscope (HIM) allows for the creation of charged defects in the MoS$_2$ lattice. The reversible drift of these locally seeded defects in the applied electric field modulates the resistance of the semiconducting channel, enabling versatile memristive functionality  on the nanoscale. We find the device can reliably retain its resistance ratios and set biases for hundreds of switching cycles at sweep frequencies of up to 2.9 V s$^{-1}$ with relatively low drain-source biases. We also demonstrate long term potentiation and depression with sharp habituation that promises application in future neuromorphic architectures. This work advances the down-scaling progress of memristive devices without sacrificing key performance parameters such as power consumption or its applicability for synaptic emulation.
\end{abstract}

In the ever-evolving field of nanoelectronics, new circuit design paradigms are constantly sought-after to improve the performance of next-generation on-chip devices. To a large extent, it is the materials used in these devices that govern their properties and potential applications. Two-dimensional (2D) layered materials such as MoS$_2$ have recently emerged as strong candidates to usher in a new age of nanoscale memories \cite{zidan2018future}. Due to enhanced electrostatic control and inter-layer van der Waals bonding, these materials could offer high-performing, flexible and transparent alternatives to silicon-based memory components \cite{chhowalla20162d,samori2016rev,jariwala2014emerging,liu2016vdw,lembke2015single,ferrari2014nanoscale,duong2017opportunities,zeng2018review}. On-chip MoS$_2$-based vertical memory cells  \cite{wo2017lowpower,vu2016floating,vu2017high,liu2018semi,wang2018robust,ge2018atomristor,ge2018radio,cheng2016ideal,huh2018synaptic} and printed lateral devices with synaptic functionalities  \cite{li2018memristors} have been prototyped in recent years.

\onecolumngrid
{\setlength\intextsep{3pt}
\begin{figure*}[h!]
\centering
\includegraphics[scale=0.15]{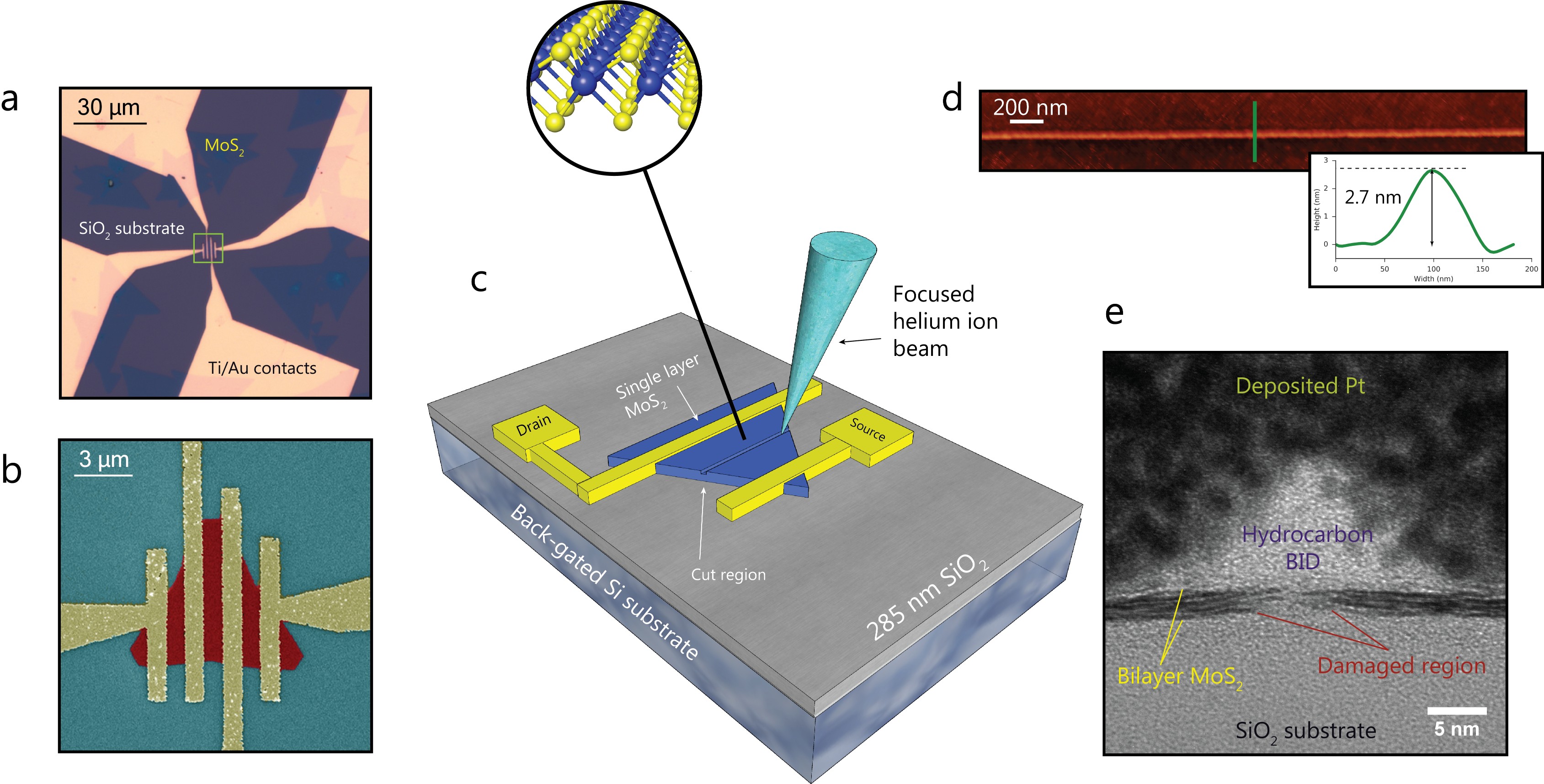}
\caption{{\footnotesize \textbf{\textit{Device fabrication and characterisation of the irradiated region}. (a) Optical micrograph of CVD MoS$_2$ flakes grown on SiO$_2$ substrates contacted with EBL-deposited Ti/Au electrodes. (b) False-colour SEM image of the region marked with the green square in (a) showing a single-crystal device contacted with multiple terminals. (c) Illustration of the irradiation strategy. (d) AFM scan of an irradiated line cut on monolayer MoS$_2$. At the delivered ion dose, the build-up of hydrocarbon beam-induced deposition is observed to average out at $\sim$ 2.7 nm along the length of the fissure. Inset: thickness profile taken from the green line marked on the AFM map. (e) Cross-sectional TEM image of a fabricated bilayer device. Underneath the hydrocarbon build-up, we observe a defect-rich region in the MoS$_2$ laterally spanning $\sim$ 6 nm.}}}
\label{fig:device}
\end{figure*}}

Single-layer polycrystalline MoS$_2$ thin films have been demonstrated to operate as gate-tunable, non-volatile and multi-terminal resistive switches; or so-called memtransistors  \cite{sangwan2015gate,sangwan2018nature}. Switching of resistance states in these devices occurs when biasing over defect-rich grain boundaries, introduced into the material during the chemical vapour deposition (CVD) process by restricting the sulfur supply. These memristive elements hold promise in mimicking neurosynaptic signalling by altering the physical state of the transistor channel through application of a large drain-source bias (> 20 V). Their unusual in-plane functionality and multi-input features integrate easily with Si-based processes; but require the implementation of a location-agnostic CVD growth process. Downscaling of this bottom-up process is ultimately limited by the dimensions of the resulting grain boundaries. The ability to locally modify the structure of transition metal dichalcogenides by scalable and industry-compatible top-down methods so that resistive switching is induced in these materials has remained a challenge.

Herein, we report the first successful fabrication of a 2D memristive switch utilising a focused helium ion beam. The structure of both CVD-grown and mechanically-exfoliated semiconducting MoS$_2$ flakes is locally altered by focused ion probe irradiations at a critical dose of 1.6 pC $\mu$m$^{-1}$. When carried out in a bisecting geometry (parallel to electrodes) this modification of the field effect transistor (FET) channel results in the creation of bipolar resistive switches. We performed microscopic and spectroscopic analyses to determine the nature of the defective interface at the irradiation site. We provide evidence that the field-driven drift of defects, believed to be sulfur vacancies (V$_{S}$) sourced from the damaged region, is the mechanism by which the different resistance states of the device are realised. In addition, we demonstrate heterosynaptic conductance modulation in multi-terminal devices and sharp habituation during long-term potentiation neural training. We highlight the pertinence of our method in producing new-generation neuromorphic circuit components, which should prove crucial for the integration of 2D materials into future memory devices.

\section*{\large{Device fabrication and defect seeding by helium ion beam}}

We employ a previously-published CVD method to grow large single-crystal triangles of monolayer MoS$_2$, as described in the \hyperref[sec:methods]{Methods} section. \hyperref[fig:device]{Figure 1a} is an optical micrograph of a typical device. The area marked with the green frame is shown magnified in the false colour scanning electron micrograph in \hyperref[fig:device]{Fig. 1b}. We note that neither the CVD-grown nor the mechanically-exfoliated MoS$_2$ samples used as starting material in this study possess visible grain boundaries, and thus their initial electrical performance is typical of pristine n-type MoS$_2$ FETs, with no hysteresis observed in the IV characteristics (see \hyperref[sec:Supp1]{Supplementary Section 1}). The irradiation strategy for inducing memristive behaviour in MoS$_2$ involves tracing out a unidirectional single pixel-wide line scan of dose 1.6 pC $\mu$m$^{-1}$ to bisect the whole FET channel, i.e. parallel to the source and drain electrodes (see geometry in \hyperref[fig:device]{Fig. 1c} and additional micrographs in \hyperref[sec:Supp1]{Supplementary Section 1}). This introduces a fissure region in the middle of the channel. The irradiation is performed once a sub-3 nm probe size is obtained by focusing elsewhere close-by on the SiO$_2$ chip (see \hyperref[sec:Supp1]{Supplementary Section 1} for determination of probe size). Exposure to the MoS$_2$ channel at this ion energy will introduce sulfur vacancies into the material and may also remove Mo atoms, as studied previously with electron energy dispersive spectroscopy  \cite{fox2015nanopatterning} and predicted by simulations  \cite{Kretschmer2018supported}. Atomic force microscopy (AFM) scans of the fissure region reveal a mound of hydrocarbons formed during beam-induced deposition (BID) capping the damaged region of MoS$_2$. For our optimal memristive dose, the height of this mound was usually found to average below 3 nm (see \hyperref[sec:Supp1]{Supplementary Section 1} for further AFM data and dependence of height on dose). We prepared a lamella of an irradiated device by focused ion beam lift-out for imaging in the transmission electron microscope (TEM). A cross-sectional TEM image of a fabricated device is presented in \hyperref[fig:device]{Fig. 1e}. We note that the extent of true lattice damage to the MoS$_2$ flake is obscured from top view by the hydrocarbon BID mound, and in actuality extends to just under 10 nanometers, where it interfaces abruptly with the pristine lattice (see plan-view TEM images in \hyperref[sec:Supp1]{Supplementary Section 1}). This is in agreement with the damage range estimated from simulations for the beam energy of 30 keV employed in our experiments  \cite{Kretschmer2018supported}. The damaged region in this work functions as a rich source of charged mobile defects. Their drift in the electric field modulates the carrier concentration in the MoS$_2$ channel, which ultimately governs the memristive behaviour of our device.

\section*{\large{Gate-tunable lateral resistive switching}}

\onecolumngrid
{\setlength\intextsep{3pt}
\begin{figure*}
\centering
\includegraphics[scale=0.4]{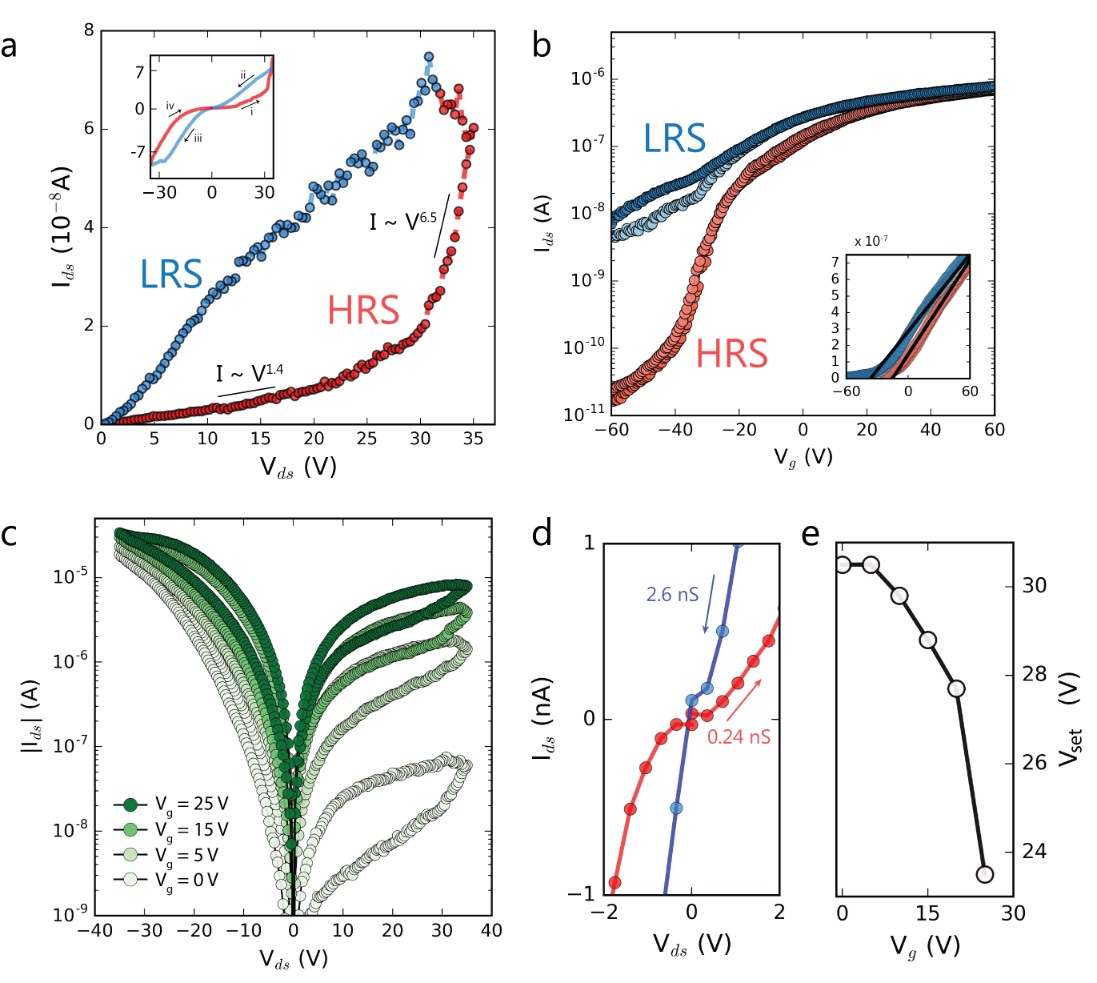}
\caption{{\footnotesize \textbf{(a) Positive-bias sweep of an irradiated monolayer memtransistor device (L = 1 $\mu$m, W = 6 $\mu$m) recorded at V$_g$ = 0 V. Inset: full range of the sweep with labelled trace directions. A sharp emergence of nonlinearity in the IV sweep marks the set point from the high resistance state (HRS) to the low resistance state (LRS). The device stays in LRS until a reset bias of negative polarity is reached on the reverse sweep. (b) Gate curves of the memtransistor taken alternately at V$_{ds}$ = 1 V in both LRS (blue) and HRS (red). Sweeps in the same state are marked by different shades of blue or red. Inset: same plot on the linear scale, with fits to the linear FET region in black extending to the threshold gate bias, V$_{th}$. (c) Gate-modulated IV sweeps of the same device. Each full cycle was repeated at the labelled V$_{g}$ values. (d) Magnified look at the zero-bias pinch-off region of the IV curve, demonstrating the conductances of the two states as V$_{ds}$ is ramped up. (e) Set bias as a function of gate voltage. Increasing positive gate fields bring about an earlier onset of resistive switching. }}}
\label{fig:switching}
\end{figure*}}

Upon the creation of our synthetic fissure with the helium beam, significant hysteresis opens up in the output curve of the device as shown in \hyperref[fig:switching]{Fig. 2a}. This device sets into the high resistance state (HRS) when first irradiated. As the drain-source bias (V$_{ds}$) is swept from 0 V to 35 V (red trace), we observe a sudden highly non-linear increase in the current, as the device switches into the low resistance state (LRS). The memtransistor remains in LRS as the bias is decreased back to 0 V (blue trace), until a change in bias polarity is introduced and the device is reset into the high resistance state. HRS persists then as the bias is swept back up to 0 V (see full sweep in inset of \hyperref[fig:switching]{Fig. 2a}). The zero-bias pinch-off of the loop, and the requirement of polarity change for resetting are hallmarks of bipolar resistive switching.

\hyperref[fig:switching]{Figure 2b} shows the transfer curves of the device in the respective resistance states. Significant shifting of the threshold voltage (V$_{th}$) is observed from -19 V in HRS to -37 V in LRS, with V$_{th}$ remaining stable through alternating gate sweeps in each state (different shades of blue and red respectively in \hyperref[fig:switching]{Fig. 2b}). High output current levels at V$_g$ = -60 V suggest the presence of surplus donors in the channel in LRS. The opening of this hysteresis is accompanied by a drop in carrier field-effect mobility relative to the unirradiated device (6.2 cm$^2$ V$^{-1}$ s$^{-1}$ before irradiation to < 1 cm$^2$ V$^{-1}$ s$^{-1}$ afterwards). No major change in mobility at V$_{g}$ = 0 V is observed between states (0.14 cm$^2$ V$^{-1}$ s$^{-1}$ and 0.16 cm$^2$ V$^{-1}$ s$^{-1}$ in HRS and LRS respectively, despite a tenfold difference in conductivity. This suggests a fundamentally similar transport mechanism in both states, with the major differentiating factor between states being the level of available donors in the channel. For gate curves of unirradiated devices and mobility extraction see \hyperref[sec:Supp2]{Supplementary Section 1}.

Application of the field effect through the back gate oxide allows for the tuning of the onset of switching, as well as the resistance ratio of the memtransistor. \hyperref[fig:switching]{Figure 2c} shows IV sweeps on a semi log scale taken at multiple gate biases. The degree of hysteresis and the resistance ratios can be manipulated by the gate bias, with values in the subthreshold regime of the transistor resulting in greater quotients (see data from other devices in \hyperref[sec:Supp2]{Supplementary Section 2}). \hyperref[fig:switching]{Figure 2d} shows a magnified look at the zero-bias pinch-off region, with disparate conductance values of both states shown on the plot. \hyperref[fig:switching]{Figure 2e} tracks the effect of the gate bias on the set voltage of the device (defined here as the bias at which a sharp non-linearity emerges in the IV curve), with increasing gate fields reducing the bias necessary to go from HRS to LRS. Our irradiation strategy makes it possible to write multiple memristive circuit elements in specified locations on a single chip. Their carrier concentrations can then be controlled at the same time by the global back-gate, opening new possibilities for novel circuit design.

We note that some devices start in LRS, exhibit multiple crossovers during the sweep and show negative differential resistance (\hyperref[sec:Supp2]{Supplementary Section 2}). However, the underlying mechanism of dopant drift, which we will discuss in detail in the spectroscopy section, remains consistent. To demonstrate the effectiveness of the helium beam in inducing resistive switching in layered materials, we repeated the experiments on mechanically-exfoliated bilayer MoS$_2$ devices and also observed similar gate-tunable memristive functionality (\hyperref[sec:Supp2]{Supplementary Section 2}); allowing our methodology to be extended to devices of larger layer thickness.

\section*{\large{Evaluation of switching parameters}}

\onecolumngrid
{\setlength\intextsep{3pt}
\begin{figure*}[h!]
\centering
\includegraphics[scale=0.45]{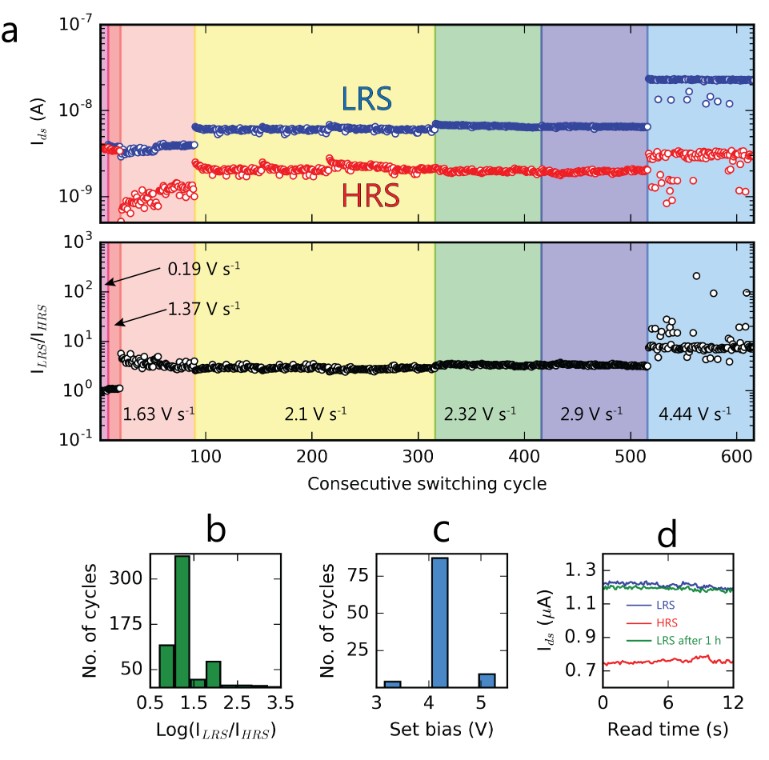}
\caption{{\footnotesize \textbf{(a) Current endurance (top) and resistance ratios (bottom) of resistance states read at V$_{ds}$ = 4 V and V$_{g}$ = 5 V across more than 600 consecutive switching cycles. Each colored portion of the plot corresponds to the sweeping frequencies used to evaluate device performance as labeled on the graph. (b) Histogram visualising the log-transformed distribution of resistance ratios from (a). (c) Set bias distribution over 100 sweeps at f$_s$ = 2.32 V s$^{-1}$ of the L = 8.5 $\mu$m device. (d) Readout of HRS and LRS currents of the device from \hyperref[fig:switching]{Fig. 2} demonstrating non-volatile retention of current levels. LRS traces were collected immediately after switching (blue) and 1 hour after cutting off the power supply (green).}}}
\label{fig:performance}
\end{figure*}}

We evaluate the switching parameters of a well-performing monolayer device when gated in the sub-threshold region (L = 8.5 $\mu$m, V$_{g}$ = 5 V, measured at $\sim$ 10 mbar) as a function of sweeping frequency, f$_s$, in \hyperref[fig:performance]{Fig. 3a}. The device was cycled through its bipolar switching IV curve between $\pm$ 20 V and the state currents were read off at V$_{ds}$ = 4 V (see evolution of epochs in \hyperref[sec:Supp3]{Supplementary Section 3}). Each colored portion of the plots indicates the f$_s$ at which the IV characteristics were swept, with f$_s$ values marked on each portion. The black points in the bottom part of the plot trace out the I$_{LRS}$/I$_{HRS}$ ratio for each cycle. The HRS and LRS currents exhibit negligible temporal variation between cycles when tested in the optimal f$_s$ range of 2.1-2.9 V s$^{-1}$. Sub-optimal frequencies on either side of this range overdamp/underdamp the ionic drift process resulting in loss of stability or no switching of resistance at all. \hyperref[fig:performance]{Figure 3b} visualises the distribution of these ratios across 616 cycles, with most of the recorded values centred between 5 and 10, which is sufficient to achieve > 80\% recognition accuracy in scaled-up neuromorphic architectures  \cite{yu2015scaling}. Set/reset bias stability is a crucial parameter that governs accurate learning in neuromorphic devices, with low stochastic variations between epochs necessary for optimal performance of resistive random access memory hardware  \cite{burr2015tolerancing,gong2018extraction}. In \hyperref[fig:performance]{Fig. 3c}, we plot the V$_{set}$ distribution extracted from 100 consecutive cycles at f$_s$ = 2.32 V s$^{-1}$. The bias is extracted based on a reached current threshold and is seen to oscillate in the initial 20 cycles (\hyperref[sec:Supp3]{Supplementary Section 3}), but settles to a steady value as the device is stabilised by more subsequent operations, with a V$_{set}$ standard deviation of $\pm$ 0.41 V on a $\pm$ 20 V sweep range. \hyperref[fig:performance]{Figure 3d} presents the retention characteristics of currents in both resistance states on the device from \hyperref[fig:performance]{Fig. 2}. The memtransistor is switched from LRS (blue) to HRS (red) and back to LRS. Power is then not supplied for 1 hour and the current in LRS is read off again (green), explicitly demonstrating the non-volatile nature of states in the memtransistor. The variability in the read current is less than 2.5\% after 1 hour. For a comparison of other parameters with the latest MoS$_2$ resistive switches from the literature, see Table S1 in \hyperref[sec:Supp3]{Supplementary Section 3}.

\section*{\large{Visualising defect drift by PL and Raman spectroscopy mapping}}
\hyperref[fig:spectro]{Figure 4} shows Raman and photoluminescence (PL) spectroscopy of a contacted monolayer CVD MoS$_2$ memtransistor (optical and scanning electron images provided in \hyperref[sec:Supp4]{Supplementary Section 4}). The channel length of the presented device is $\approx$ 1 $\mu$m, extending from 0.5 $\mu$m to 1.5 $\mu$m on the x-axes in the figures. Raman and PL maps were  collected in each resistance state (see \hyperref[sec:methods]{Methods} for details of mapping parameters). We remark that our devices were mapped and electrically switched in ambient conditions. To visualise the mapping results, the E$^{'}$ Raman mode at 386 cm$^{-1}$ was summed over a spectral window of $\pm$ 5 cm$^{-1}$ on either side of the peak while the B exciton emission at 2.01 eV was summed over a spectral window of 0.1 eV. \hyperref[fig:spectro]{Figures 4a,b} show line profile intensities of the E$^{'}$ Raman mode averaged across the device channel, alternating between LRS (a) and HRS (b). The source and drain electrodes are shown in gold on either side of the MoS$_2$ channel, with the red arrows indicating direction of mobile defect drift under the applied bias. We distinguish the drain and source sides of the device which are separated by a He$^+$-created fissure of our standard dose of 1.6 pC $\mu$m$^{-1}$. For the effect of different doses on Raman/PL see \hyperref[sec:Supp4]{Supplementary Section 4}. The location of the He$^+$ cut is indicated in the figures by the dotted black lines.

In LRS, the spatial distribution of the Raman intensity on either side of the black line is roughly equal, as represented by the uniform blue color in (c). The red areas represent low signal regions from the MoS$_2$ hidden by the electrodes.  In HRS (d), the intensity is observed to drop off strongly on the drain side and is distributed non-uniformly in the material (for the averaged Raman spectra on each side of the channel see \hyperref[sec:Supp4]{Supplementary Section 4}). \hyperref[fig:spectro]{Figures 4e-h} show analogous PL profiles and maps. As noted in the Raman results, the emission on the source side does not change significantly by switching from LRS to HRS. Once again, the drain side emission declines in intensity; with particular quenching of the B exciton (factor of $\sim$7) in the PL maps (for the averaged PL spectra on each side of the channel see \hyperref[sec:Supp4]{Supplementary Section 4}). The spatial uniformity of the emission on the drain-side is disrupted in HRS, with most of the signal from the top and bottom regions of the channel quenching drastically between states.

We observe no significant shifts or broadening of the E$^{'}$ and A$_{1}^{'}$ Raman peaks on the drain-side, despite a large drop in intensity. The separation of the in-plane E$^{'}$ (386 cm$^{-1}$) and out-of-plane A$_{1}^{'}$ (405 cm$^{-1}$) first-order modes is $\sim$20 cm$^{-1}$, consistent with single-layer MoS$_2$  \cite{lee2010anomalous} as we switch between states; maintaining good integrity of the crystal structure. The drop in Raman intensity suggests a large decrease in electron-phonon coupling as defects migrate from the fissure region into the unirradiated lattice. PL emission from the direct-recombination A exciton and spin-orbit split B exciton is centred at 1.89 eV and 2.01 eV respectively throughout the switching with no shifts. This range is usually reported for typical monolayer MoS$_2$  \cite{splendiani2010emerging,chow2015defect}, also confirming the relative structural integrity of the material in both LRS and HRS.

\onecolumngrid
{\setlength\intextsep{3pt}
\begin{figure*}
\centering
\includegraphics[scale=0.35]{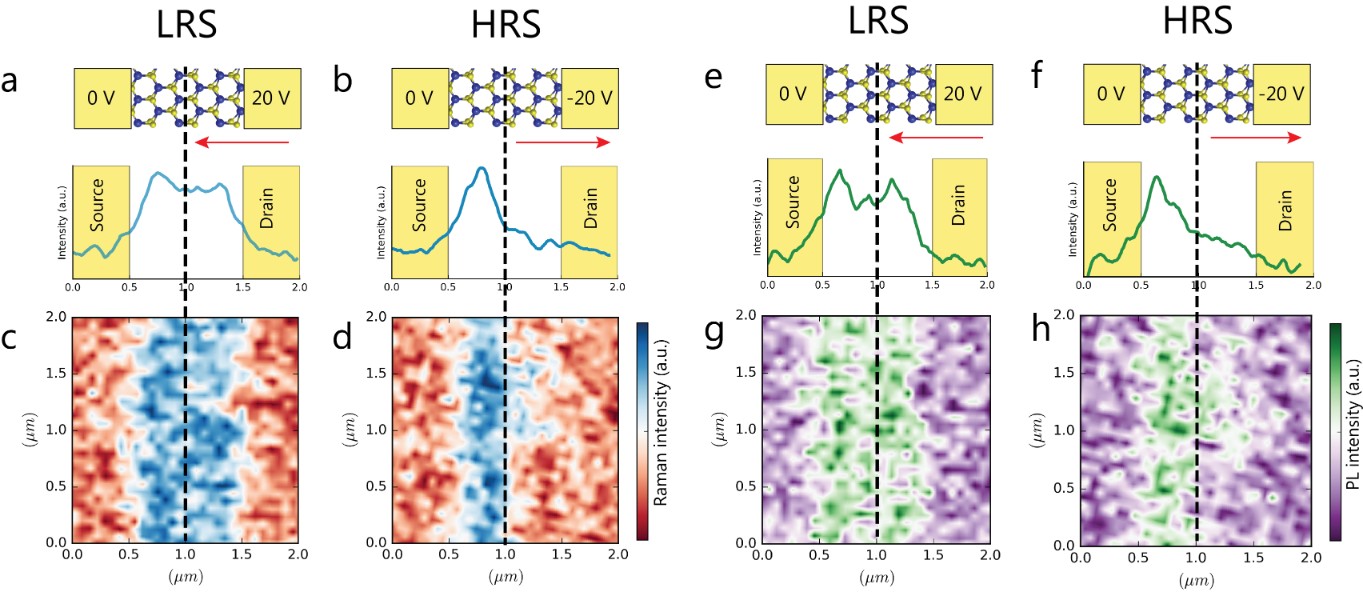}
\caption{{\footnotesize \textbf{\textit{Raman and PL spectroscopic characterisation of the device in different resistance states} (a,b) Line profiles of Raman intensities across the He$^+$ ion cut. The red arrow marks the direction of mobile dopant drift when the applied bias is as written on electrodes in the sketch above (not to scale). (c,d) Spatially resolved Raman intensity maps of the E$^{'}$ mode, normalised to the colour bar on the right. Severe quenching of signal is observed on the drain-side in HRS. (e-h) Analogous B exciton photoluminescence data from the same device. PL data also shows considerable loss of intensity on the drain-side of the fissure after switching into HRS.}}}
\label{fig:spectro}
\end{figure*}}

Based on our results, we propose that when switching the device into HRS, the defects on the drain-side are forced to migrate towards the drain electrode from the fissure. Their depletion from the channel attenuates B exciton intensity which is usually associated with defect concentration  \cite{chow2015defect,kaupmees2018trions}. Helium ion irradiation above doses of > 1 pC $\mu$m$^{-1}$ results in V$_S$ concentrations \cite{fox2015nanopatterning} induced by beam sputtering that far exceed sub-stoichiometries achieved naturally by CVD growth \cite{kim2014influence}.  As sulfur vacancies are defects with the lowest formation energy in MoS$_2$  \cite{komsa2015native}, are stable on substrates when charged \cite{urasaki2018first}, and also have a lowered energy barrier for diffusion when in clusters  \cite{komsa2013from,le2014catalytic} and when charged  \cite{sensoy2017strain}, we propose that they are the majority species responsible for resistance switching in our devices. However, computational works have challenged the role of V$_{S}$ in donating electrons to the MoS$_2$ lattice  \cite{komsa2015native,shang2018eliminate} so we cannot rule out Mo interstitials or antisite defects, inevitably introduced by the helium beam, as also contributing to the mechanism. The approximate concentration of helium-generated V$_{S}$ in LRS on the drain-side of the fissure is at most $\sim$ 4.2 $\cdot$ 10$^{14}$ V$_{S}$ cm$^{-2}$ (see \hyperref[sec:Supp4]{Supplementary Section 4} for calculation). From our transfer curves in \hyperref[fig:switching]{Fig. 2b}, we can estimate the difference in available carrier concentration at the same gate bias between LRS and HRS from the V$_{th}$ shift to be $\sim $1.6 $\cdot$ 10$^{12}$ cm$^{-2}$. This shows that the available amount of potential mobile donors provided by the ion beam is comfortably sufficient to account for the difference detected by gating between LRS and HRS. We note that a downshift of E$^{'}$ is expected if the V$_S$ concentration is higher than 1\%  \cite{parkin2016raman}. As the difference in V$_S$ concentration between LRS and HRS in our case is less than 1\%, we do not detect any peak shifts in the Raman spectra.

The electro-chemical metallisation process occurring at the MoS$_2$/contact interface  \cite{sangwan2018nature} effectively strips V$_{S}$ of their ability to donate free charge to the FET channel in HRS. The optical properties of the source-side of our device in this experiment were not affected. We propose that defects from the drain-side are not able to readily pass the beam-created barrier in this monolayer device. Our spectroscopic observations are in line with transfer curves obtained in HRS and LRS (\hyperref[fig:switching]{Fig. 2b}), showing a reduced degree of n-type doping in HRS, as donor defects are depleted at the drain electrode interface.

\onecolumngrid
{\setlength\intextsep{3pt}
\begin{figure*}
\centering
\includegraphics[scale=0.35]{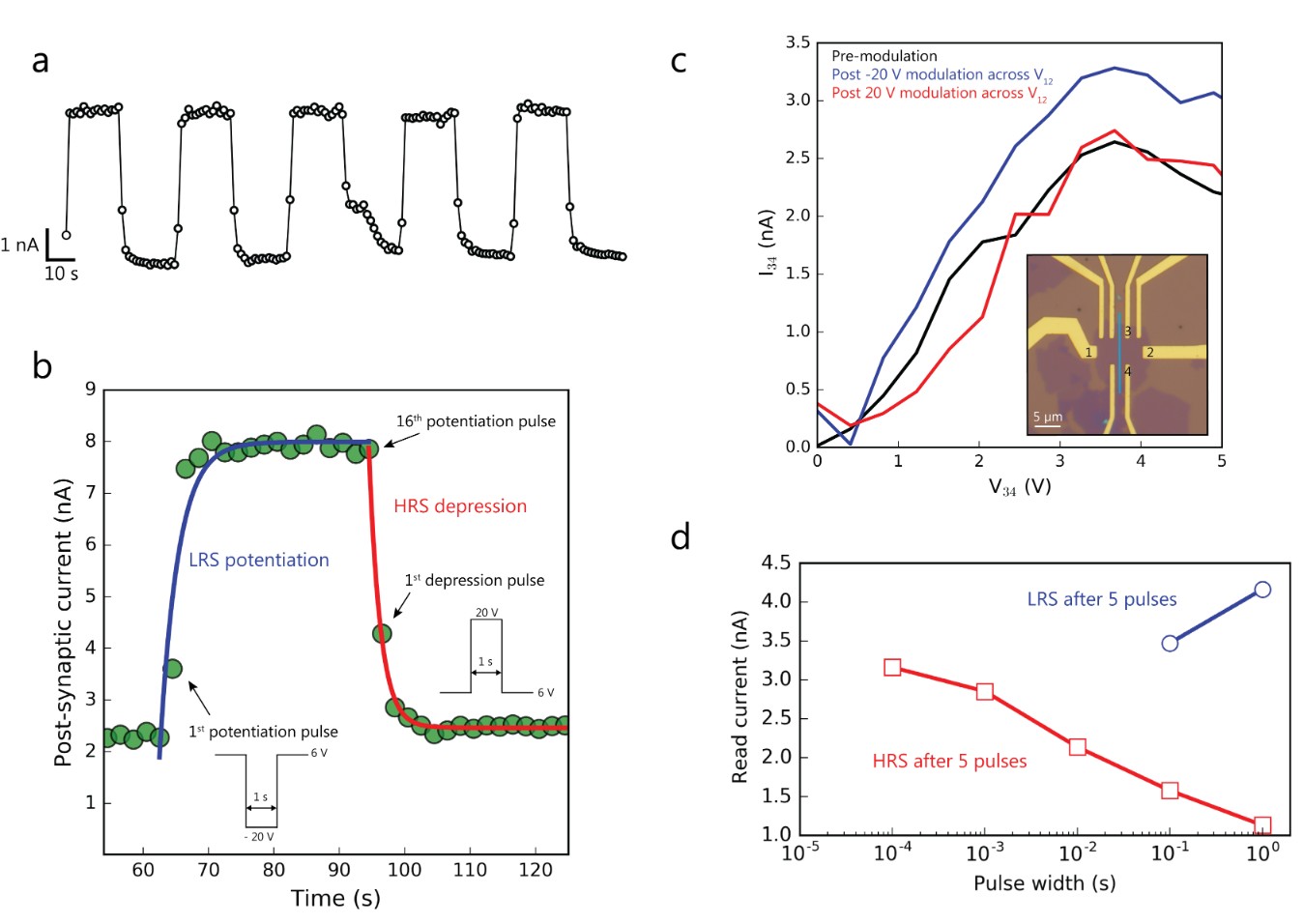}
\caption{{\footnotesize \textbf{(a) Trace of post-synaptic current recorded during 5 sequential LTP/LTD training cycles. Trains of 16 pulses were applied past the reset voltage into LRS and set voltage into HRS. (b) Closer look at the second potentiation/depression cycle from (a). Both the rise and decay readouts are fitted well with simple exponentials and habituate to saturated resistance values in each state within < 5 pulses.  (c) Memtransistor cross-terminal modulation enabled by lateral drift of defects sourced from the cut region. Inset is an optical micrograph of the device with the particular terminals labeled and the cut region marked in blue. (d) Dependence of read current on the width of the pre-synaptic learning pulse. Shorter timescales inhibit complete habituation to LRS below 0.1 s at the applied read voltage of 5 V. }}}
\label{fig:synaptic}
\end{figure*}}

\section*{\large{Synaptic potentiation and cross-terminal conductance modulation}}

The success of two-dimensional materials in future neuromorphic architectures will largely depend on their ease of implementation with on-dielectric integrated circuits, as well as their ability to mimic synaptic processes with good accuracy. Training pulses which simulate neuronal input activity were used to demonstrate the neuromorphic applicability of our devices. Consecutive cycles of long term potentiation (LTP) and depression (LTD) were recorded (\hyperref[fig:synaptic]{Fig. 5a}) by stressing the device into HRS or LRS respectively with a train of 16 evenly-spaced action potential pulses. The post-excitation current was read out at V$_{ds}$ = 6 V following each pulse, and is fitted well with a sharp exponential rise/decay function (\hyperref[fig:synaptic]{Fig. 5b}). The memtransistor was always back-gated at V$_g$ = 5 V throughout this testing. We note that the lowest power consumption (disregarding gating) in standby is $P = I_{HRS} \cdot V_{read} = 16$ nW, which compares favourably at the same bias with the previously-reported polycrystalline films and multi-layer devices ($\mu$W range in HRS for both \cite{sangwan2018nature,li2018memristors}). With optimisations, such as reduction of channel length and contact area, our devices should start approaching sub-pW levels demonstrated by some vertically-stacked 2D devices  \cite{ge2018radio,shi2018synaptic}.

As demonstrated recently by Sangwan et al. \cite{sangwan2018nature}, planar MoS$_2$ memtransistors enable emulation of heterosynaptic plasticity through seamless incorporation of multi-terminal architectures. \hyperref[fig:synaptic]{Figure 5c} presents low-bias IV traces recorded between terminals 3 and 4 of the device in the inset image. By stressing the channel across the He$^+$ ion fissure (between contacts 1 and 2), defects can migrate between the cut region and electrode 2. This modulates the conductance of the MoS$_2$ channel on the right-hand side of the artificial defect source, significantly altering the output current between perpendicular terminals 3 and 4. This also directly demonstrates that the resistance of the channel is a function of mobile defect concentration. The output current of the 3-4 junction is modulated multifold only by stressing the perpendicular 1-2 junction, even below its V$_{set}$.

The logic and switching applications of any memristive device will ultimately be limited by the frequency at which states can be accessed and read out. Functionalities such as pattern recognition rely on effective thresholding of output currents  \cite{prezioso2015training,sheridan2017sparse} and smearing of state conductivities at high binary switching frequencies will lower the tolerance required for multi-level operations  \cite{bessonov2015memcap}. Our devices maintain state distinguishability above switching rates of 10 Hz on an 8.5 $\mu$m channel containing a single sub-10 nm fissure. \hyperref[fig:synaptic]{Figure 5d} tracks the dependence of read currents on the width of the applied 5-pulse train used to stress the device into the different resistance states. HRS can be reliably accessed down below the millisecond range, while LRS down to 0.1 s (for binary switching capabilities of the device see \hyperref[sec:Supp5]{Supplementary Section 5}).

\section*{Conclusions}

In summary, we have successfully demonstrated an experimental approach to locally creating gate-tunable memristive circuit elements in atomically-thin materials. The advantages of employing a helium ion microscope include high material selectivity for sputtering, remarkable nanoscale precision and scalability of fabrication. Additionally, the instrument is strongly compatible with established industrial lithographic processes. In contrast to growing polycrystalline films, our strategy allows for site-specific inclusion of memristive elements in large-area monolayer circuits, as well as post-metallization alterations. With stable neuromorphic functionality demonstrated, we advance the potential of two-dimensional semiconductors for inclusion in future revolutionary nanoelectronic devices.

\clearpage

\section*{Methods}
\label{sec:methods}

\textbf{Device fabrication}

\textit{In Trinity College Dublin}:
Monolayer MoS$_2$ sheets were grown by proximate sulfurisation of MoO$_3$ precursors on marked 285 nm SiO$_2$/Si chips inside a chemical vapour deposition furnace using a technique described previously in ref \cite{obrien2014proximity}. Briefly; sulfur powder was heated at 120$^{\circ}$C and the resulting vapour carried downstream by Ar gas to a hot zone kept at 750$^{\circ}$C. The vapour was flown through a microreactor containing MoO$_3$ film on a seed substrate. The film was sulfurised and deposited on target substrates facing the seed substrate for 20 minutes. The samples were then annealed in Ar for 20 minutes at 750$^{\circ}$C. Suitable MoS$_2$ triangles were identified using an \textit{Olympus} optical microscope (50$\times$ lens objective). Electrode patterns were designed on \textit{Raith Nanopatterning} software and defined by electron beam lithography (\textit{Zeiss Supra SEM}), utilising PMMA A3 resist exposed at a dose of 200 $\mu$C cm$^{-2}$ followed by development for 45 seconds in MIBK:IPA (1:3). Metallization was carried out in an electron beam-evaporator (\textit{Temescal FC-2000}) to deposit small-area Ti/Au (5/40 nm) pads, followed by lift-off in acetone overnight at room temperature. Bilayer samples were mechanically exfoliated from bulk MoS$_2$ (\textit{SPI Supplies}) with adhesive tape and identified using optical contrast and AFM.

\textit{In Peking University}:
Transported specimens grown in Trinity College were marked out using a \textit{Nikon} optical microscope (100$\times$ lens objective). Electrode patterns were designed on \textit{Raith Nanopatterning} software and defined by electron beam lithography (\textit{FEI Strata DB 235 SEM}), utilising PMMA A9 resist exposed at a dose of 340 $\mu$C cm$^{-2}$ followed by development for 30 seconds in MIBK:IPA (1:3). Metallization was carried out in an electron beam-evaporator (\textit{Detech DE400}) to deposit large-area Ti/Au (10/80 nm) pads, followed by lift-off in warm acetone.

\noindent \textbf{Helium ion microscope irradiations}

\textit{In Trinity College Dublin and Peking University}:
The fabricated devices were inserted into the \textit{Zeiss Nanofab} helium ion microscope 24 hours before irradiation to allow for outgassing at a base pressure of $\sim$ 10$^{-7}$mbar. The milling patterns were designed on the \textit{NanoPatterningVisualisationEngine} software and consisted of a single-direction pixel-wide scan traced in a single sweep with no retracing. The delivered ion linear dose to induce memristive switching was $\sim$1.6 pC $\mu$m$^{-1}$ at a beam current of 1.5 pA, aperture of 10 $\mu$m, with the probe size evaluated at < 3 nm using the GaussFit module in ImageJ software (determination of probe size in \hyperref[sec:Supp1]{Supplementary Section 1}).

\noindent \textbf{Electrical Testing}

\textit{In Trinity College Dublin}: The samples were tested in the vacuum chamber of a customised scanning electron microscope (\textit{Zeiss EVO}), at a base pressure of $\sim$ 10$^{-5}$ mbar. \textit{Imina miBot} piezoelectric tungsten probes were used to contact the EBL-deposited pads as source and drain terminals. The data were obtained using a semiconductor analyzer (\textit{Agilent B2912A}) interfaced with \textit{Keysight} software. The samples were back-gated with copper tape underneath the 285 nm SiO$_2$/Si chip, and a third probe was used as the gate terminal.

\textit{In Peking University}: The samples were wire-bonded using aluminium leads to large-area Ti/Au pads, and globally back-gated through a silver paint-connected terminal underneath the highly-doped Si substrate. Endurance and potentiation measurements were carried out in an \textit{Oxford Instruments} cryostat with no applied magnetic field, pressure of $\approx$ 10 mbar and temperature of 280 K. The data were obtained using a semiconductor analyzer (\textit{Agilent B2912A}) interfaced with \textit{LabVIEW} software.

\noindent \textbf{Raman and photoluminescence spectroscopy}

Spectroscopic mapping was done in a \textit{WITec Alpha 300R} system in ambient conditions. The device was switched between resistance states \textit{ex-situ} between each map acquisition, and the maps were acquired within minutes of switching states. The excitation wavelength was 532 nm, with a diffraction grating of 1800 grooves mm$^{-1}$ and a 100$\times$ objective lens (NA = 0.95, spot size $\sim$ 0.3 $\mu$m). Mapping was carried out at low incident powers (< 100 $\mu$W) to avoid altering the sample with the laser, with the beam refocused manually every time between each scan to ensure accurate comparison. The 4 $\mu$m$^2$ areas scanned in 60 lines in \hyperref[fig:spectro]{Fig. 4} were sampled at a resolution of 60 px per line, with integration time of 0.6 s.

\noindent \textbf{Focused ion beam lamella preparation, scanning electron microscopy, transmission electron microscopy and atomic force microscopy}

The helium-treated lamella was prepared from a mechanically-exfoliated bilayer MoS$_2$ sample (\textit{SPI Supplies}), cleaved onto a 285 nm-SiO$_2$/Si substrate with adhesive tape. Once irradiated in the HIM, the sample was transferred to a focused ion beam system (\textit{Zeiss Auriga}), where a platinum layer was deposited to protect the flake. A gallium beam (30 keV) was used to cut the lamella for lift-out and welding to a TEM grid made of copper. The sample was then thinned to an electron-transparent width with a 900 eV argon beam (\textit{Fischione 1040 Nanomill}).

SEM images of the devices pre- and post-irradiation were taken in \textit{Zeiss Supra} and \textit{Zeiss Ultra} systems, with chamber pressures at $\sim$ 2 $\cdot$ 10$^{-6}$ mbar.

TEM was performed in an \textit{FEI Titan 80-300} microscope operated at a chamber pressure of $\sim$ 4 $\cdot$ 10$^{-7}$ mbar. The beam energy used for imaging was 300 keV. Irradiated samples for plan-view imaging were transferred from the HIM to the TEM immediately after exposure. Freestanding samples were obtained by covering the on-chip samples with PMMA, dissolving the underlying SiO$_2$ in NaOH at 80$^{\circ}$C and transferring onto copper TEM grids; followed by dissolution and washing in acetone/IPA.

Devices were scanned in an \textit{Oxford Asylum} atomic force microscope with 140 kHz cantilevers operated in ambient conditions.

\section*{Acknowledgements}

We thank Ms. Rongrong Li for assistance with the helium ion microscope at Peking University. We acknowledge members of staff at the Advanced Microscopy Laboratory and CRANN clean room, Trinity College Dublin for their continued technical support. We thank Mr. Emmet Sheerin for assistance with switching devices during the Raman and PL experiments. We acknowledge support from the following funding bodies: National Natural Science Foundation of China (Nos. 61825401 and 11774004), Thousand Young Talents Program of China, Leverhulme Trust International Networks Grant (PicoFIB), Science Foundation Ireland (Nos: 11/PI/1105, 12/TIDA/I2433, 07/SK/I1220a and 08/CE/I1432), the Irish Research Council (Nos: GOIPG/2014/972 and EPSPG/2011/239), European Research Council (No: Advanced Grant 321160).

\section*{Author contributions}

$^{\dagger}$J.J. and D.K. contributed equally to this project.

D.K. performed electrical measurements on samples of different layer thickness and irradiation dose, as well as atomic force microscopy and scanning electron microscopy. CVD growth of MoS$_2$ monolayers was carried out by C.P.C. Mechanically-exfoliated devices were prepared by J.J. and D.K. Raman and PL spectroscopy was carried out by C.P.C. and analysed by P.M.. J.J and D.K. carried out EBL to fabricate the FET devices. H.S. assisted with fabrication and wirebonding of devices tested in Peking University (PKU). D.K., J.J. and P.M. carried out the HIM irradiations in Trinity College Dublin (TCD) while J.J and Y.Z. conducted the HIM exposures in PKU. J.J carried out the electrical tests (endurance and potentiation) with assistance from H.S and Y.Z in PKU. P.M. performed FIB processing of irradiated devices and carried out TEM of the cross-sectioned lamellae with assistance from C.D.. D.S.F. carried out helium exposures and TEM imaging of the plan-view irradiated devices after J.J. transferred the samples onto TEM grids. Z.L. oversaw the electrical characterisation work in PKU, while R.Z. and J.X. facilitated microscopy experiments in PKU. N.M and G.S.D. oversaw the material growth process and spectroscopic experiments in TCD. J.J.B. and H.Z. conceived the study and supervised the project. The manuscript was written by J.J., D.K. and P.M. All authors agreed with the final version of the paper.

\section*{Additional information}

Supplementary information containing additional experimental data for each of the sections discussed in this manuscript is attached at the end of this document.

Correspondence and requests for materials should be addressed to Hongzhou Zhang.

\section*{Competing financial interests}
The authors declare no competing financial interests.

\bibliographystyle{plain}

\appendix
\clearpage

\onecolumngrid

\section*{Supplementary Information}

\subsection*{Supplementary Section 1}
\label{sec:Supp1}

\subsubsection*{Determination of He$^+$ ion probe size}

{\setlength\intextsep{3pt}
\begin{figure*}[h!]
\centering
\includegraphics[scale=0.3]{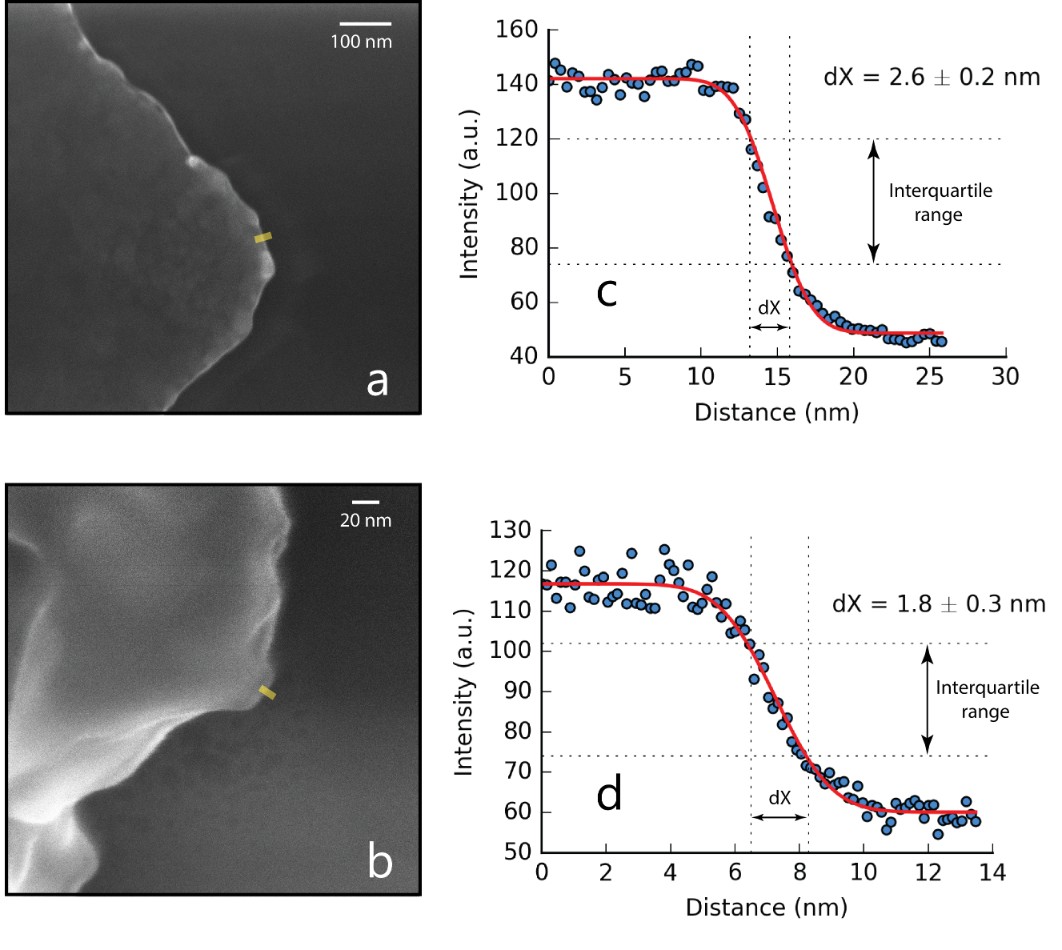}
\caption{{\footnotesize \textbf{\textit{Evaluation of He$^+$ probe size before irradiation on device presented in \hyperref[fig:performance]{Fig. 3a}.} (a,b) HIM images of the edges of deposited gold marks near an irradiation site. (c,d) Probe size estimation using the 25/75\% profile method from line profiles seen on the HIM images in yellow. }}}
\end{figure*}

Prior to exposure of the MoS$_2$ channel, the probe size of the helium ion beam was established from images of nearby gold marks elsewhere on the substrate (within tens of microns away from the target device). As gold has a much higher secondary electron emission intensity than the underlying oxide substrate, it is assumed that the edge of the gold mark is ideally sharp.

\textit{ImageJ} software was used to fit the intensity drop-off over the edge of the gold mark on a 2048 pixel$^2$ micrograph taken with a scan dwell time of 5 $\mu$s. The equivalent full-width-at-half-maximum of the fit is taken to be the probe size as dictated by the Gaussian spatial distribution of the ions in the impinging beam.

\subsubsection*{Extended AFM data}
{\setlength\intextsep{3pt}
\begin{figure*}[h!]
\centering
\includegraphics[scale=0.3]{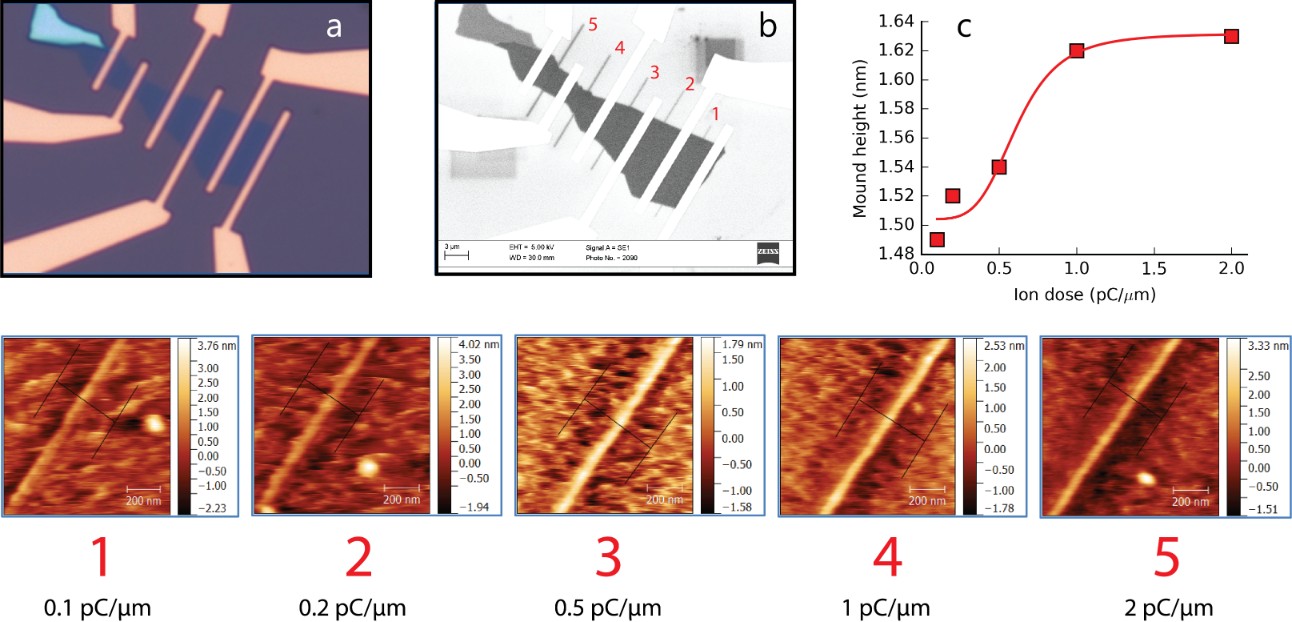}
\caption{{\footnotesize \textbf{\textit{Dependence of hydrocarbon BID mound height on delivered ion dose} (a,b) Optical and SEM images of a monolayer device irradiated at different helium ion doses. The 1-5 labels correspond to the AFM maps presented below. (c) Extracted mound heights from the regions marked in AFM maps at each respective dose. The solid line is a saturating fit scaling as $\propto \displaystyle \frac{1}{1 + d^{p}}$ where $d$ is the dose and $p$ is a fitting exponent.}}}
\end{figure*}

{\setlength\intextsep{3pt}
\begin{figure*}[h!]
\centering
\includegraphics[scale=0.3]{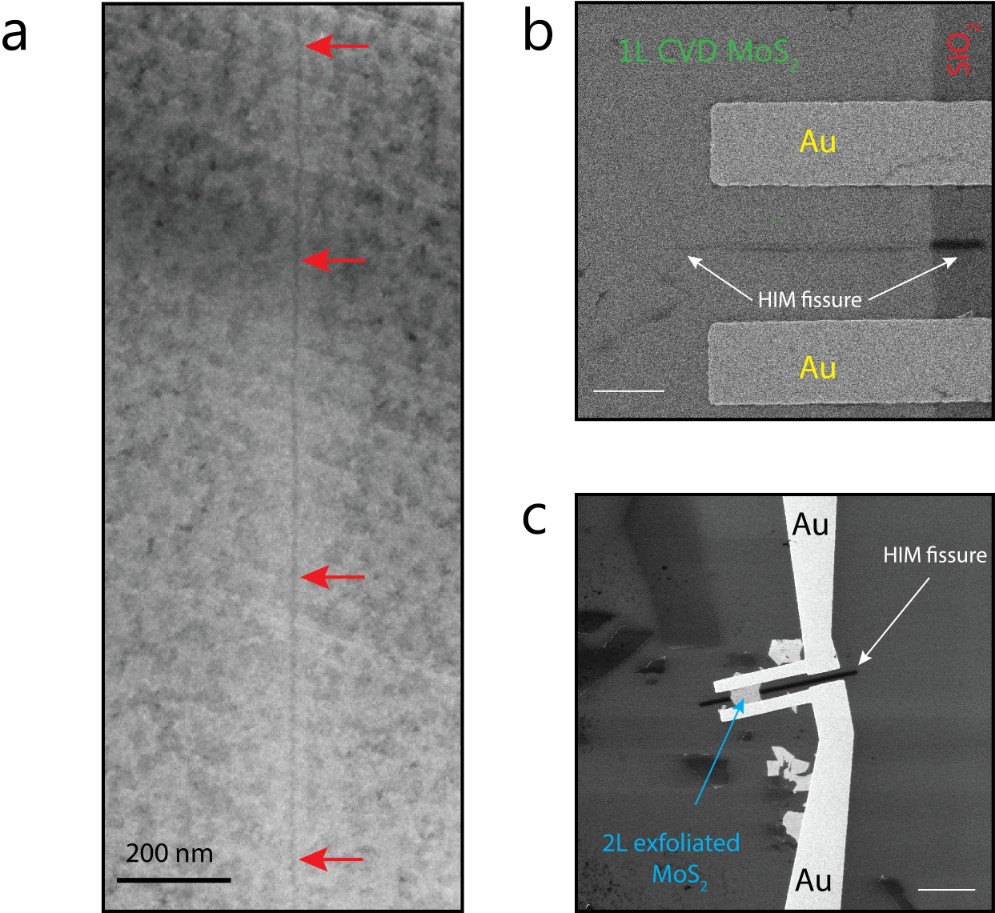}
\caption{{\footnotesize \textbf{\textit{SEM and HIM images of irradiated devices.} (a) High-magnification image of the fissure region with minimal beam-induced deposition build-up on the surface. The ion-damaged region is faintly seen running from top to bottom of the image; its right edge is marked by the red arrows. (b) HIM image of an irradiated 1L CVD device, showing darker contrast of the cut on the SiO$_2$ substrate due to differing secondary electron yield. Scale bar is 3 $\mu$m. (c) HIM image of a 2L MoS$_2$ memtransistor, showing the fissure region extending out onto the substrate on both sides of the device to ensure complete channel bisection. Scale bar is 5 $\mu$m.}}}
\end{figure*}}

{\setlength\intextsep{3pt}
\begin{figure*}[h!]
\centering
\includegraphics[scale=0.4]{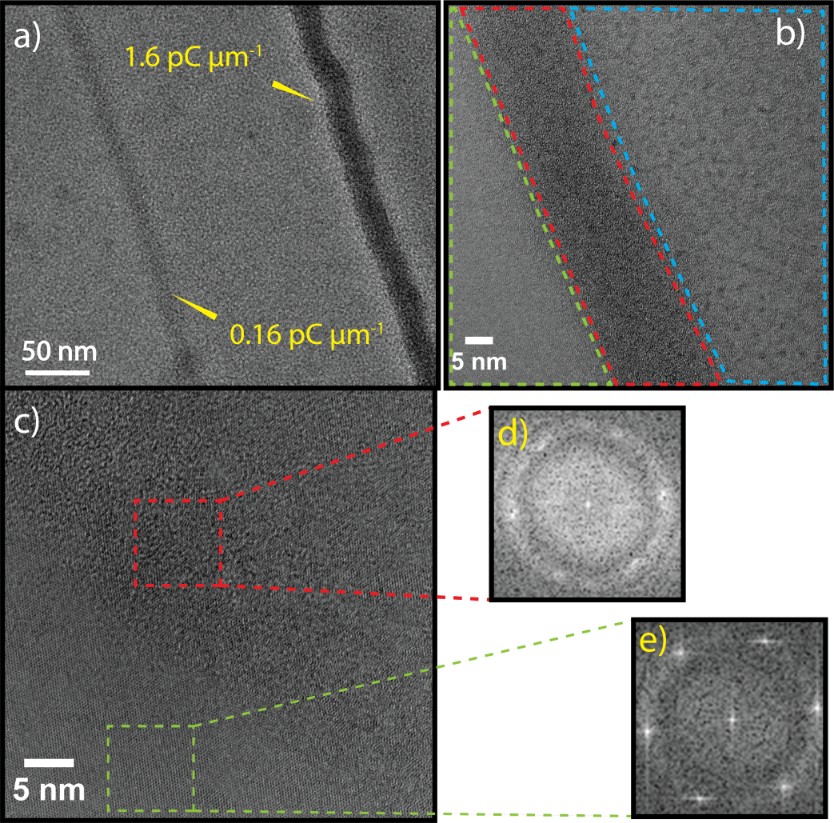}
\caption{{\footnotesize \textbf{\textit{Plan-view TEM images of cuts performed on suspended MoS$_2$ at ion doses of 0.16 and 1.6 pC $\mu$m$^{-1}$.} (a) Line scans side-by-side on the same sample, showing visible difference in phase contrast due to larger BID buildup on the higher dose cut. (b) Magnified look at the 1.6 pC $\mu$m$^{-1}$ region in red. Areas immediately adjacent to the cut show different levels of defectivity (blue higher, green lower), suggesting that the ion beam may preferentially induce switching to one side of the cut if the probe is stigmated or the sample is not perfectly flat. (c) Closer look at the interface between the green and red regions. The boundary between the irradiated region and pristine MoS$_2$ extends only a few atomic rows. (d) and (e) show Fast Fourier Transforms from the marked regions revealing a more crystalline lattice in the unirradiated region. }}}
\end{figure*}}
\clearpage

{\setlength\intextsep{3pt}
\begin{figure*}[h!]
\centering
\includegraphics[scale=0.3]{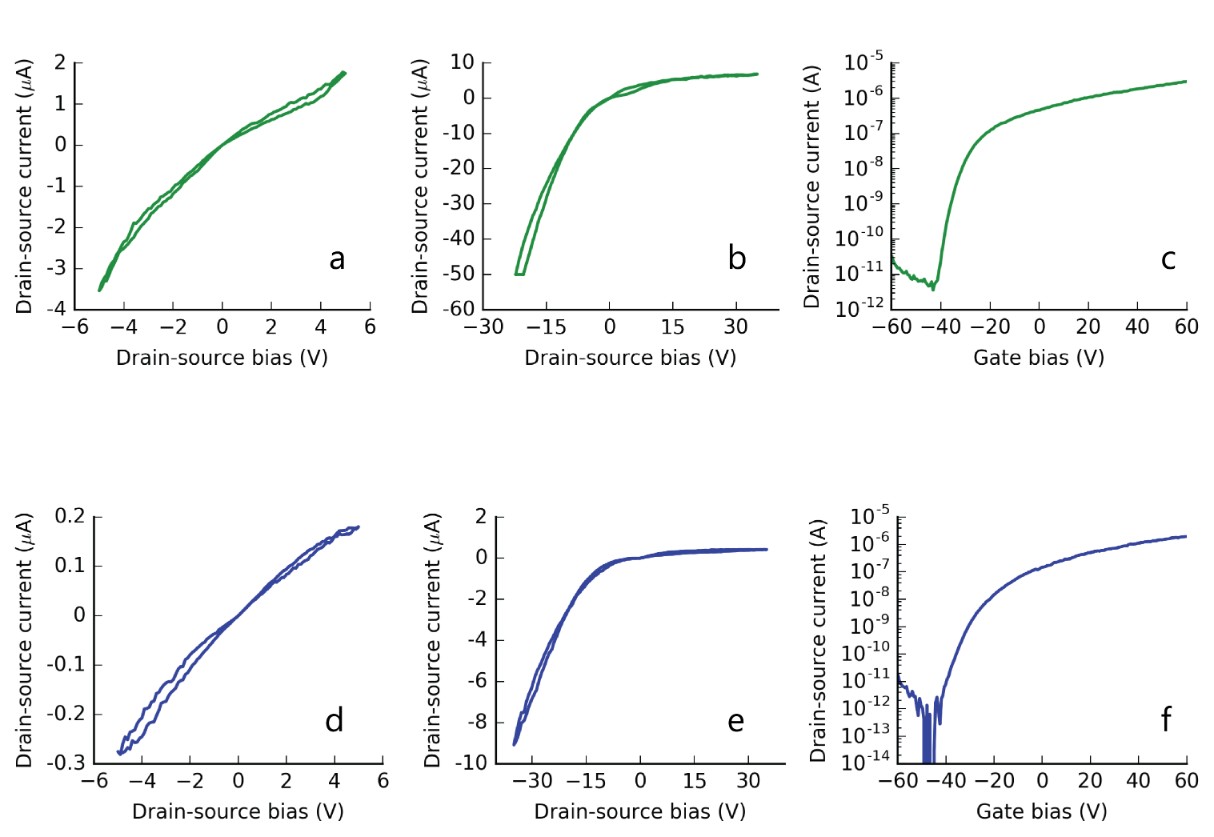}
\caption{{\footnotesize \textbf{\textit{Typical IV and gate curves of unirradiated CVD devices.} (a,b,c) IV curves between $\pm$5 V, $\pm$35 V and gate curve respectively for the same device. (d,e,f) IV curves between $\pm$5 V, $\pm$35 V and gate curve respectively for another device on the same chip. In both cases, no significant hysteresis opens up in the output characteristics even at large drain-source biases. Transfer curves are typical of n-type MoS$_2$ with field-effect mobilities in the linear region not exceeding 10 cm$^2$ V$^{-1}$ s$^{-1}$. }}}
\end{figure*}

The carrier field-effect mobility, $\mu_\mathrm{fe}$ was extracted in the linear region of the transfer curve (V$_g$ = 0 V) using the expression:
\begin{equation}
  \mu_\mathrm{fe} = \frac{L}{V_{ds} C_{ox} W} \cdot \frac{\partial I_{ds}}{\partial V_{g}}
\end{equation}
where $L$, $W$ are the length and width of the conducting channel, $V_{ds}$ is the drain-source bias, $C_{ox}$ is the gate oxide capacitance evaluated at 13.9 nF cm$^{-2}$, $I_{ds}$ is the drain-source current evaluated at the gate bias, $V_g$.

This same formula was applied to evaluate the mobilities in \hyperref[fig:switching]{Fig. 2b} of the main text.

\clearpage
\subsection*{Supplementary Section 2}
\label{sec:Supp2}

{\setlength\intextsep{3pt}
\begin{figure*}[h!]
\centering
\includegraphics[scale=0.4]{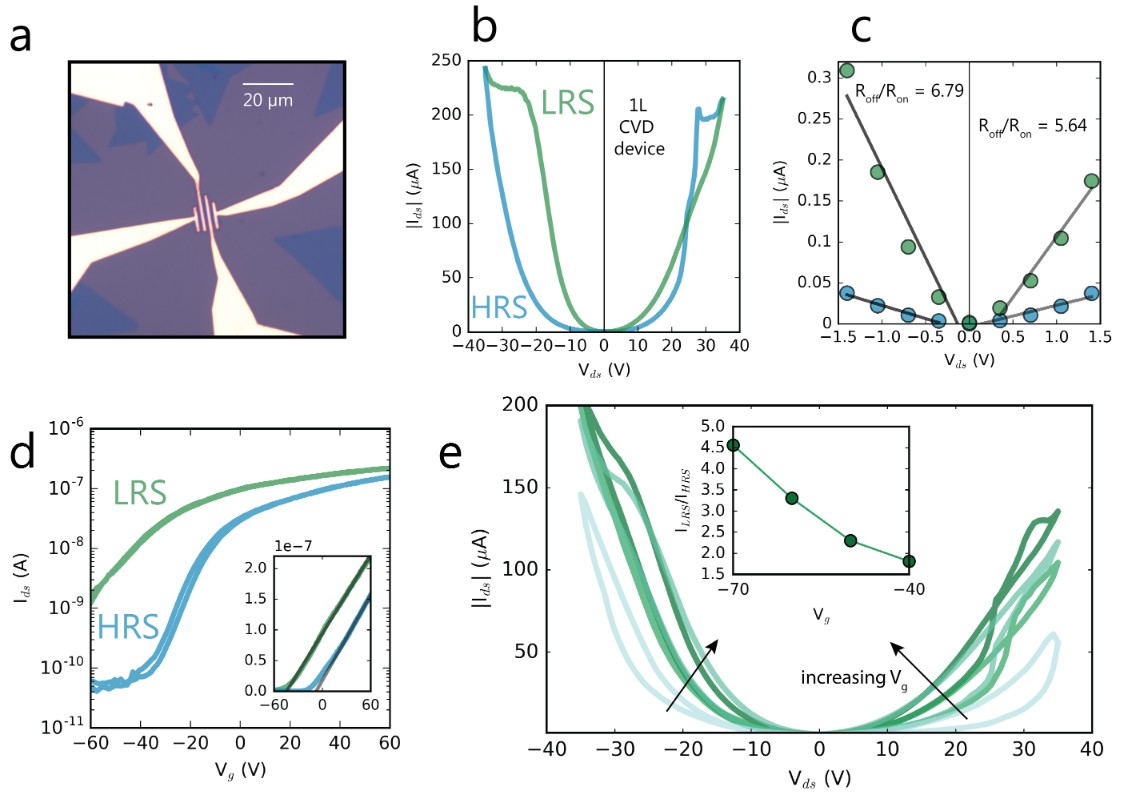}
\caption{{\footnotesize \textbf{\textit{Examples of switching on another 1L CVD device.} (a) Optical image of the device. (b) IV curves showing distinct resistance states with an additional crossover in the positive section of the hysteretic loop. (c) Resistance ratios fitted in the low bias regime. (d) Gate sweeps of the device demonstrating LRS threshold bias shifting to negative gate voltages. Inset is the same plot on the linear scale. (e) Gate-tunable looping behaviour of the device. The black arrow marks the direction of increasing V$_g$. Inset: Dependence of resistance ratios on the applied gate bias. The ratio increases with increasing negative polarity.}}}
\end{figure*}

{\setlength\intextsep{3pt}
\begin{figure*}[h!]
\centering
\includegraphics[scale=0.6]{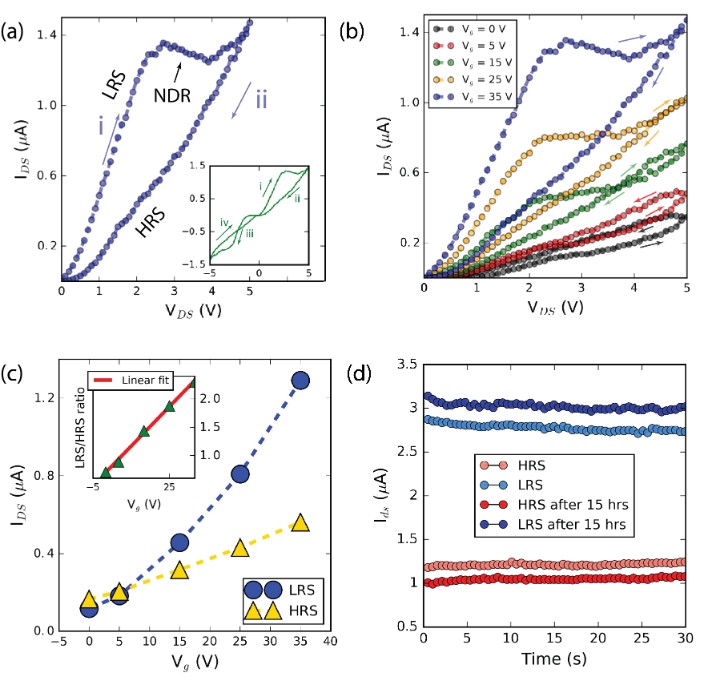}
\caption{{\footnotesize \textbf{\textit{Switching behaviour in mechanically-exfoliated bilayer MoS$_2$.} (a) IV curve of bilayer device, starting in LRS and demonstrating negative differential resistance before switching to HRS. Inset is the full range curve, with an additional resistance crossover on the negative side of the sweep. (b) Gate field tunability of the hysteretic loop; higher positive biases reduce the resistance ratio. (c) Resistance state read currents at different values of the back-gate bias extracted at the largest discrepancy point between states in the IV loop. Inset shows the linear scaling of this absolute largest resistance ratio with increasing gate bias; which follows from the scaling of V$_{set}$ with V$_g$. (d) Retention times of resistance states in the bilayer device, showing current levels in each state after cutting off power supply for 15 hours.}}}
\end{figure*}

\clearpage
\subsection*{Supplementary Section 3}
\label{sec:Supp3}

{\setlength\intextsep{3pt}
\begin{figure*}[h!]
\centering
\includegraphics[scale=0.4]{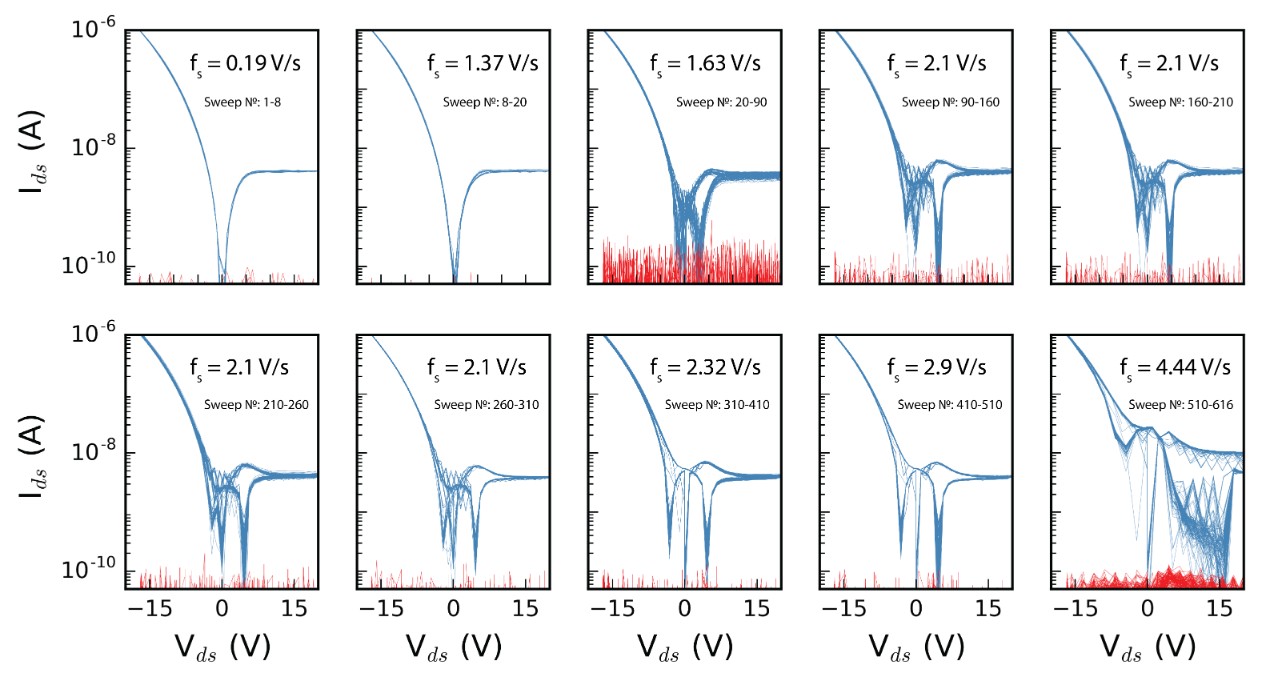}
\caption{{\footnotesize \textbf{\textit{Consecutive switching cycles of the device used to generate data in \hyperref[fig:performance]{Fig. 3a}.} The graphs are organised in order of consecutive testing, with the sweep frequency and cycle numbers labelled on each plot. Blue traces are the drain-source current recorded on sweeps of V$_{ds}$ = $\pm$20 V. The red trace is the recorded gate leakage current for each cycle, showing negligible leakage close to the noise floor of the sourcemeter.}}}
\end{figure*}

{\setlength\intextsep{3pt}
\begin{figure*}[h!]
\centering
\includegraphics[scale=0.3]{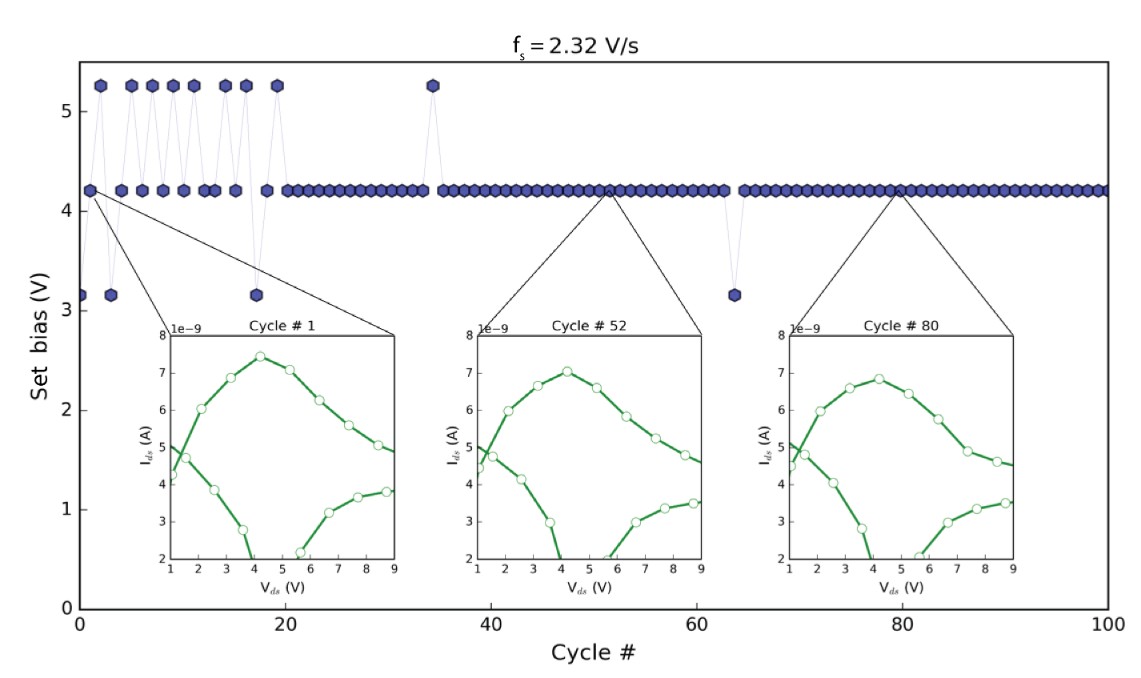}
\caption{{\footnotesize \textbf{\textit{Set bias stability over 100 switching cycles at f$_s$ = 2.32 V s$^{-1}$ from \hyperref[fig:performance]{Figs. 3a,c}.} The shape of the hysteretic loop varies around the V$_{set}$ as the device is stabilised by consecutive switching cycles, and the maximum current level drops slightly.}}}
\end{figure*}

\clearpage
\begin{center}
\begin{adjustbox}{width = {20cm},angle={90}}

\begin{tabular}{c|c|c|c|c|c|c|c|c}
\textbf{Reference}               & \textbf{Summary}                                               & \textbf{Geometry}             & \textbf{Reported V$_{set}$}  & \textbf{Resistance ratio}            & \textbf{Retention time}                  & \textbf{Endurance}                             & \textbf{Switching timeframe}              & \textbf{Power consumption}          \\ \hline
{\color[HTML]{3166FF} This work} & {\color[HTML]{3166FF} HIM-fabricated}                          & {\color[HTML]{3166FF} Planar} & {\color[HTML]{3166FF} 3.5 V} & {\color[HTML]{3166FF} $\sim$ 10} & {\color[HTML]{3166FF} \textgreater 1 hr} & {\color[HTML]{3166FF} \textgreater 616 cycles} & {\color[HTML]{3166FF} \textgreater 0.1 s} & {\color[HTML]{3166FF} $\sim$ 16 nW} \\
\cite{ge2018atomristor}                       & 1L TMD between Cr/Au pads                                      & Vertical                      & 1 V                          & \textgreater 10$^4$              & \textgreater 1 week                      & 150 cycles                                     & 1 s                                       & $\sim$ 10 nW                        \\
\cite{wang2018robust}                           & Stacked graphene/MoS$_{2-x}$O$_x$/graphene                     & Vertical                      & 3.5 V                        & $\sim$ 10                        & $\sim$ 1 week                            & \textgreater 10$^7$ cycles                     & 100 ns                                    & \textgreater 1 $\mu$W               \\
\cite{li2018memristors}                       & Mechanically printed 15 nm MoS$_2$ with Au/Ti pads             & Planar                        & 15 V                         & \textless 10                     & \textgreater 30 seconds                  & \textgreater 5 cycles                          & 2 ms                                      & \textless 0.05 W                    \\
\cite{sangwan2018nature}                       & Polycrystalline large-area monolayer CVD films with Ti/Au pads & Planar                        & 80 V                         & $\sim$ 100                       & 24 hours                                 & 475 cycles                                     & 1 ms                                      & 50 pW - 1 mW                        \\
\cite{cheng2016ideal}                       & (92\% 1T) MoS$_2$ sandwiched between silver electrodes         & Vertical                      & 66 mV                        & $\sim$ 1000                      & N/A                                      & 1000 cycles                                    & N/A                                       & 0.1 mW                              \\
\cite{sangwan2015gate}                     & CVD-grown monolayers with individual grain boundary            & Planar                        & 3.5 - 8.4 V                  & 1000                             & 120 s                                    & 15 cycles                                      & N/A                                       & \textgreater 1 $\mu$W               \\
\cite{bessonov2015memcap}                        & Annealed MoS$_2$/MoO$_x$ matrix between silver electrodes      & Vertical                      & 150 mV                       & 10$^4$                           & \textgreater 1000 s                      & \textgreater 10$^4$ cycles                     & 5 ms                                      & \textless 1 $\mu$W                  \\
\cite{ge2018radio}                      & 1L MoS$_2$ between Cr/Au pads                                  & Vertical                      & 1 V                          & 10$^4$                           & \textgreater 10$^4$ s                    & 20 cycles                                      & \textless 30 ns                           & \textless 1 $\mu$W
\end{tabular}
\end{adjustbox}
\end{center}

\textbf{Table S1:} Summary of various device parameters from relevant MoS$_2$ memory devices in the literature. The presently-discussed work is highlighted in blue.

\clearpage
\subsection*{Supplementary Section 4}
\label{sec:Supp4}

{\setlength\intextsep{3pt}
\begin{figure*}[h!]
\centering
\includegraphics[scale=0.5]{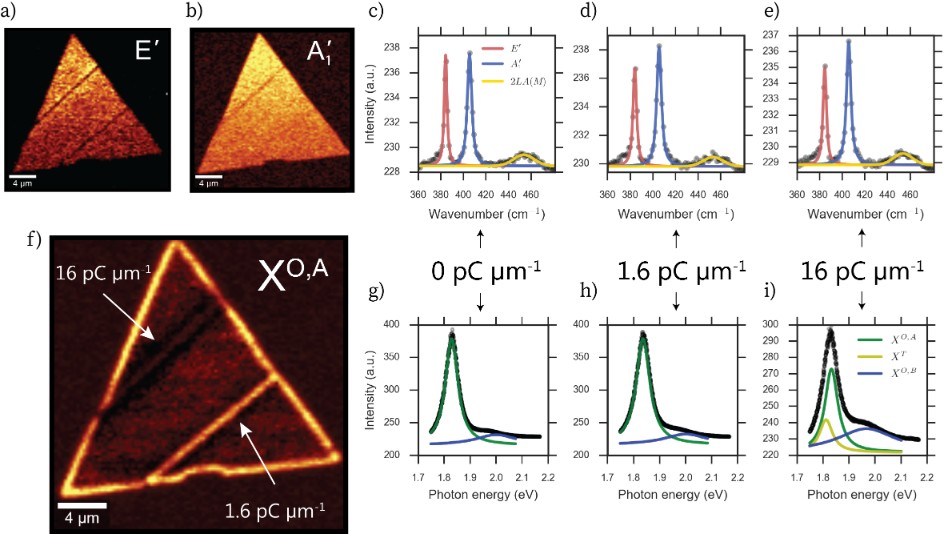}
\caption{{\footnotesize \textbf{\textit{Raman and PL spectroscopic characterisation of helium ion fissures created at doses of 1.6 pC $\mu$m$^{-1}$ and 16 pC $\mu$m$^{-1}$.} (a,b) Raman maps filtered for the E$^{'}$ and A$_{1}^{'}$ monolayer modes, showing an effect of peak shifting (red-shift of 0.2 cm$^{-1}$ and further 0.6 cm$^{-1}$ after 1.6 pC $\mu$m$^{-1}$ and 16 pC $\mu$m$^{-1}$ respectively for E$^{'}$; and blue-shift of 0.6 cm$^{-1}$ and further red-shift of 0.6 cm$^{-1}$ after 1.6 pC $\mu$m$^{-1}$ and 16 pC $\mu$m$^{-1}$ respectively for A$_{1}^{'}$). No obvious emergence of defect-related peaks is observed in the spectra. Raman quenching of the E$^{'}$ mode due to ion irradiation of the material in the tested area is stronger than that of the A$_{1}^{'}$ mode, suggesting possible exposed MoS$_2$ edges in the region (see similarity between helium fissures and edges of the as-grown flake in both the Raman and PL). Spectra in (c-e) are extracted from a pristine middle region (0 pC $\mu$m$^{-1}$) and the fissure regions from edge-to-edge of the flake parallel to the fissure. Different ion dose spectra are marked with arrows in the figure. No significant peak shifting is observed even at the high 16 pC $\mu$m$^{-1}$ dose, but the intensity is reduced. (f) PL map of the A exciton intensity, showing enhanced emission at the site of the memristive dose (h) due to an increased concentration of available adsorption sites \cite{tongay2013broad,tongay2013defects, nan2014enhancement}. The higher dose destroys the material significantly, however, quenching the PL emission relative to (g) and (h) and introducing a defect-associated trion peak \cite{lin2014screening} as seen in yellow on the spectrum in (i).}}}
\end{figure*}}

{\setlength\intextsep{3pt}
\begin{figure*}[h!]
\centering
\includegraphics[scale=0.25]{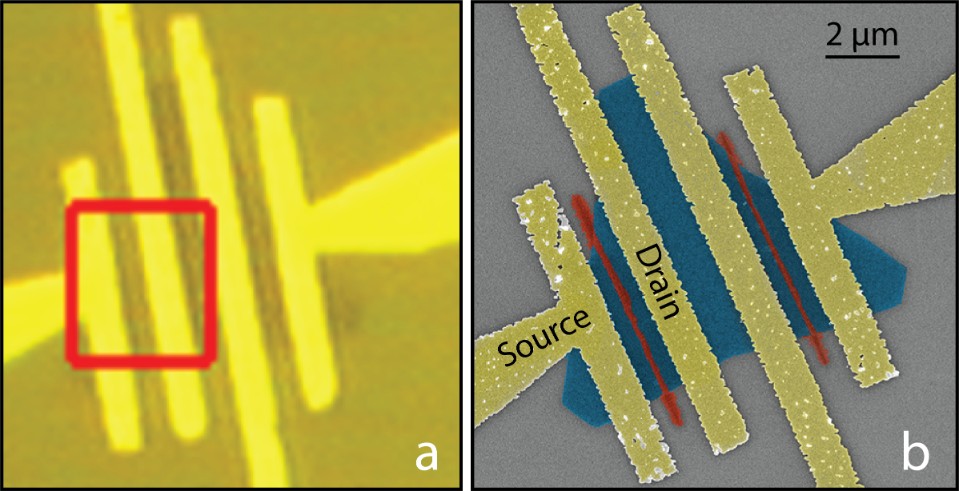}
\caption{{\footnotesize \textbf{\textit{Device used for Raman and PL mapping in \hyperref[fig:spectro]{Fig. 4}.} (a) Optical micrograph (100$\times$ lens), showing the square area which was mapped outlined in red. (b) False-color SEM image of the same device post-irradiation, with the helium ion fissures marked in red and the source and drain electrodes labeled as used in the experiment.}}}
\end{figure*}}
\clearpage
{\setlength\intextsep{3pt}
\begin{figure*}[h!]
\centering
\includegraphics[scale=0.4]{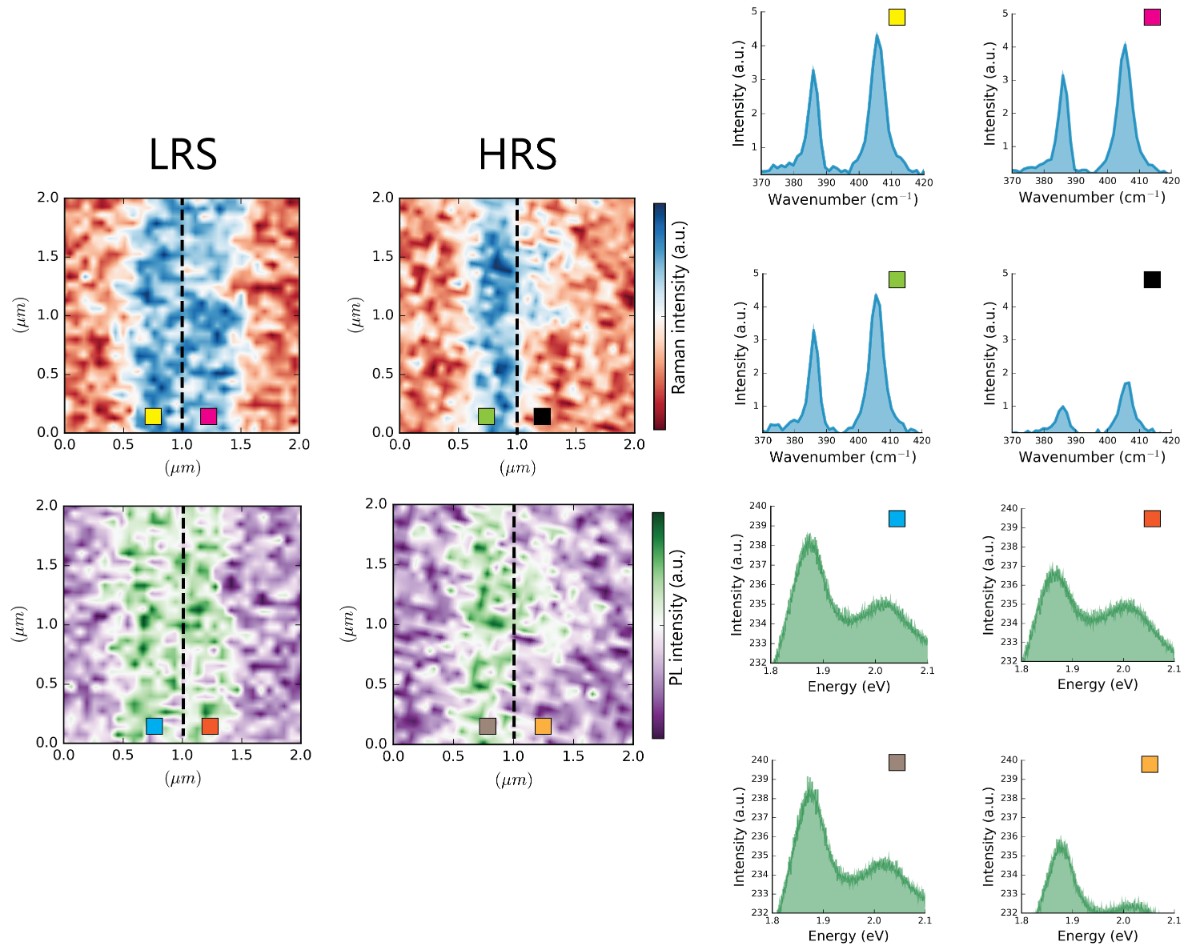}
\caption{{\footnotesize \textbf{\textit{Raman and PL spectra extracted from maps in \hyperref[fig:spectro]{Fig. 4}.} Each coloured square indicates the fissure side from which the corresponding spectrum was averaged (900 spectra per each average in the region spanning 0-0.5 $\mu$m or 1-1.5 $\mu$m on the x-axis, and 0-2 $\mu$m on the y-axis). Quenching of both Raman and PL intensities is visible on the drain-side (right hand side) in HRS, as seen on the spectra marked with the black and orange squares. }}}
\end{figure*}}
\clearpage

\begin{table}[h!]
  \centering
\begin{tabular}{|c|c|c|c|c|}
\hline
Resistance state                                & \multicolumn{2}{c|}{\textbf{LRS}} & \multicolumn{2}{c|}{\textbf{HRS}} \\ \hline
Area relative to fissure                     & \textit{Source}        & \textit{Drain}      & \textit{Source}        & \textit{Drain}      \\ \hline
$\omega_{A^{'}_{1}}$ (cm$^{-1}$)                & 405.8       & 405.4      & 405.6       & 405.8      \\ \hline
I$_{A^{'}_{1}}$ (a.u.)                          & 3.89        & 3.69       & 4.12        & 1.6        \\ \hline
$\Gamma_{A^{'}_{1}}$ (cm$^{-1}$)                & 5.1         & 5.3        & 5.3         & 5.1        \\ \hline
$\omega_{E^{'}}$ (cm$^{-1}$) & 386.2       & 386.2      & 386.1       & 385.8      \\ \hline
$\Gamma_{E^{'}}$ (cm$^{-1}$) & 3.2         & 3.2        & 3.3         & 4          \\ \hline
I$_{E^{'}}$ (a.u.)        & 2.67        & 2.72       & 2.78        & 0.65       \\ \hline
A$_{1}^{'}$/E$^{'}$ intensity ratio                             & 1.46        & 1.36       & 1.48        & 2.13       \\ \hline
\end{tabular}
\end{table}

\textbf{Table S2:} Summary of fitting parameters to spectra presented in \hyperref[fig:spectro]{Fig. 4}. No major peak shifts are noted, but the intensity of both peaks decreases on the right side of the channel in HRS. That region is also the only one where the E$^{'}$ peak is more intense than the A$_{1}^{'}$ peak, suggesting reduced out-of-plane vibration intensities due to lattice rearrangement.
\clearpage

\subsection*{Estimation of available mobile vacancy concentration}

{\setlength\intextsep{3pt}
\begin{figure*}[h!]
\centering
\includegraphics[scale=0.5]{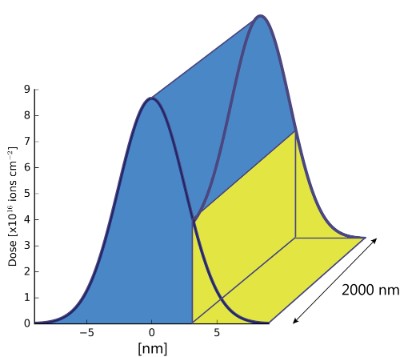}
\caption{{\footnotesize \textbf{\textit{Sketch of simulated dose delivery along a line scan of probe size 6 nm, as based on the TEM cross-section from \hyperref[fig:spectro]{Fig. 1e}.}}}}
\end{figure*}}

The above figure is a simulated areal helium ion dose distribution for a line scan with overlapping beam probes (spacing 1 nm). The Gaussian distribution of the ion dose is blurred by a 6 nm probe size, as evaluated from the visibly damaged region in the TEM cross-section in \hyperref[fig:spectro]{Fig. 1e}. The centre of the fissure is assumed to be at 0 nm on the x-axis in the figure.

We assume that any ions delivered to the MoS$_2$ lattice integrated under the blue region cannot cross or leave the fissure, effectively not contributing to the creation of mobile V$_S$ species. The region shaded in yellow marks the concentration of delivered ions that will be able to sputter sulfur atoms and leave behind potential mobile vacancies on the drain-side of the device.

The integrated areal dose in the yellow portion of the curve was calculated to be 6.4 $\times$ 10$^{16}$ ions cm$^{-2}$. As the sputter yield of S per helium ion is $\approx$ 0.007 \cite{maguire2018defect}, the final higher bound for the available V$_S$ concentration sourced from the yellow strip is 4.2 $\times$ 10$^{14}$ V$_S$ cm$^{-2}$. This is comfortably above the experimentally-detected carrier concentration difference between LRS and HRS of 1.57 $\times$ 10$^{12}$ cm$^{-2}$, suggesting that even more vacancies get trapped in the fissure region; possibly due to backscattered and secondary atom sputtering events extending the fissure region beyond the assumed probe size confinement.


\clearpage
\subsection*{Supplementary Section 5}
\label{sec:Supp5}

{\setlength\intextsep{3pt}
\begin{figure*}[h!]
\centering
\includegraphics[scale=0.5]{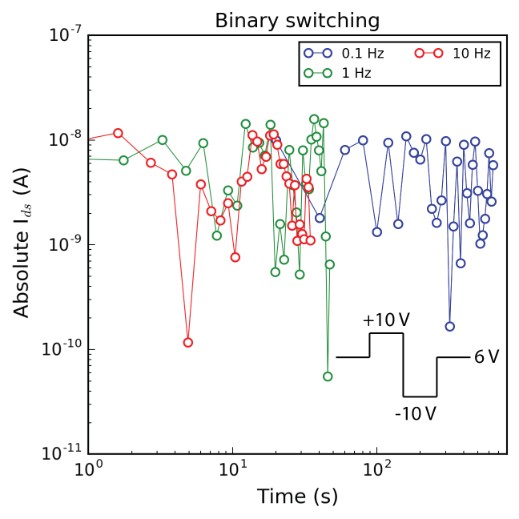}
\caption{{\footnotesize \textbf{\textit{On-off switching between binary states as a function of pulse width.} Alternating pulses of $\pm$10 V were used to continually force the device into LRS or HRS with state readout performed at V$_{ds}$ = 6 V. As the width of the pulses narrows, a frequency-dependent smearing of states emerges above 1 Hz, where the on/off states are hard to distinguish (green \& red traces) and the readout currents are no longer consistent. Defect migration is hence rate-limited and the operational frequency needs to be well-paired with device geometry to achieve optimal performance.}}}
\end{figure*}}

\end{document}